\documentclass{aa}

\usepackage[varg]{txfonts}
\usepackage{graphicx}
\usepackage{array, booktabs, makecell}
\usepackage{siunitx}
\usepackage[version=4]{mhchem}
\usepackage{textcomp, gensymb}
\usepackage[export]{adjustbox}
\usepackage{amsmath}
\usepackage{amsfonts}
\usepackage{comment}
\usepackage{multicol}
\usepackage[utf8]{inputenc}
\usepackage[english]{babel}
\usepackage{natbib}
\bibpunct{(}{)}{;}{a}{}{,}
\usepackage[hidelinks,colorlinks=true,linkcolor=blue,citecolor=blue]{hyperref}
\usepackage[dvipsnames]{xcolor}

\usepackage{orcidlink}

\begin{document}

\title{Millimeter VLBI constraints on the central magnetic field and symmetric jet production in the twin-jet galaxy NGC\,1052} 

\author{L. Ricci \orcidlink{0000-0002-4175-3194}\inst{1,2},
A.-K. Baczko \orcidlink{0000-0002-4175-3194}\inst{3,2},
M. Kadler
\orcidlink{0000-0001-5606-6154}\inst{1},
C. M. Fromm
\orcidlink{0000-0002-1827-1656}\inst{1},
A. Saiz-Pérez
\orcidlink{0000-0002-4175-3194}\inst{1},
E. Ros
\orcidlink{0000-0001-9503-4892}\inst{2},
S. del Palacio
\orcidlink{0000-0002-5761-2417}\inst{3},
T. P. Krichbaum
\orcidlink{0000-0002-4892-9586}\inst{2},
M. Perucho
\orcidlink{0000-0003-2784-0379}\inst{4,5},
G. F. Paraschos
\orcidlink{0000-0001-6757-3098}\inst{2}
}

\institute{
\inst{1} Julius-Maximilians-Universit{\"a}t W{\"u}rzburg, Fakult{\"a}t für Physik und 
Astronomie, Institut für Theoretische Physik und Astrophysik, 
Lehrstuhl für Astronomie, Emil-Fischer-Str. 31, D-97074 W{\"u}rzburg, 
Germany \\
\inst{2} Max-Planck-Institut f{\"u}r Radioastronomie, Auf dem H{\"u}gel 69, D-53121 Bonn, Germany \\
\inst{3} Department of Space, Earth and Environment, Chalmers University of Technology, SE-41296 Gothenburg, Sweden \\
\inst{4} Departament d’Astronomia i Astrofísica, Universitat de València, C/ Dr. Moliner, 50, E-46100, Burjassot, València, Spain \\
\inst{5} Observatori Astronòmic, Universitat de València, C/ Catedràtic José Beltrán 2, E-46980, Paterna, València, Spain \\
}

\date{Received / Accepted}

  \abstract
{}
{This paper investigates the symmetry and magnetic field properties of the jets in the radio galaxy NGC\,1052, with particular attention to the impact of the ionized torus that surrounds the central region on the emitted radiation.}
{Our study is based on three new 43\,GHz Very-Long-Baseline Interferometry (VLBI) observations and one 86\,GHz observation conducted between April 2021 and April 2022. 
We derive key jet parameters, such as speed, width, and flux density for both jets at the two frequencies and compare them with those obtained from previous VLBI campaigns. 
Additionally, we present the first (43-86)\,GHz spectral index image of NGC\,1052, which is crucial to assess the role of the torus at high frequencies.
Finally, we leverage the derived observational parameters to constrain the magnetic field strength and configuration in the launched jets.}
{We observe variability in the jet morphology at 43\,GHz across the three epochs, which can be associated with the propagation of jet knots launched from the nuclear region. 
The stacked 43\,GHz image reveals that the western and receding jet is approximately three times fainter than its eastern (approaching) counterpart in the sub-mas region. 
This asymmetry, together with the (43-86)\,GHz spectral index map, suggests that free--free absorption may affect the 43\,GHz emission.
On the contrary, the jets appear highly symmetric at 86\,GHz. 
From the stacked images at 43\,GHz and 86\,GHz, we extract the jet width, which is consistent with previous VLBI studies and supports the presence of a parabolic jet profile on very compact scales. 
Overall, our results suggest that the jets are intrinsically launched symmetrically, and that the observed time-dependent asymmetries may result from free--free absorption by the torus and the downstream propagation of jet components, a scenario supported by previous theoretical studies.
Finally, we estimate the magnetic field strength along both jets, finding good agreement with earlier works, and discuss the possible presence of a magnetically arrested disk (MAD) in the nuclear region of NGC\,1052.}
   {}
   
\titlerunning{mm-VLBI constraints on the central magnetic field and symmetric jet production in the twin-jet galaxy NGC\,1052}
\authorrunning{L. Ricci et al.}
\maketitle

\section{Introduction} \label{sec:introduction}

Active galactic nuclei (AGN) are known to produce powerful relativistic jets that can propagate on scales up to thousands of kpc \citep[see, for example][and references therein]{Blandford2019}. 
One key parameter regulating jet launching and propagation is the magnetic field permeating the entire system, from the central region down to the extended jets.
The relativistic jets are expected to be launched in the form of Poynting-flux dominated outflows, with initial magnetic field strengths in the range of hundreds to thousands of Gauss, which decrease down to milliGauss on parsec scales \citep[see, for example,][]{O'Sullivan2009}.
On (sub-)parsec scales, magnetic fields play a crucial role in collimating and accelerating the outflows, with magnetic energy being converted into kinetic energy of the bulk flow until equipartition is reached \citep{Vlahakis2003_a, Vlahakis2003_b, Vlahakis2004, Komissarov2007, Komissarov2012, Ricci2024}.
On compact scales, multiple studies highlighted the presence of helical magnetic field configurations \citep[see, for example,][]{Hovatta2014, Gabuzda2018}, pointing towards both the toroidal and poloidal field components to play an important role on such scales.
However, whether the magnetic field structure is universal in AGN jets or varies across different types of sources is an ongoing investigation \citep[see, for example,][and references therein]{Boccardi2017, Hodge2018}.

Concerning the event horizon scales, when considering their magnetization, hot and radiatively inefficient accretion disks can be divided into the two main models of standard and normal evolution disks \citep[SANE,][]{Narayan_2012} and magnetically arrested disks \citep[MAD,][]{Bisno_1974, Bisno_1976, Narayan_2003, Tchek2011}.
The first represents the low-magnetization scenario in which particles lose angular momentum due to the magnetorotational instability.
On the contrary, the MADs are highly magnetized disks permeated by strong poloidal magnetic fields that can saturate and disrupt the accretion flow, forming a magnetosphere surrounding the central engine.
The MAD model is gaining increasing visibility in recent years thanks to its ability to successfully model the launching of powerful jets from hot accretion disks \citep[see, for example,][]{Narayan2022}, and is currently proposed for an increasing number of low-luminosity AGN \citep{Zamaninasab2014, EHT2021, Yuan2022, Ricci2022}.

In this context, the mentioned mechanisms are expected to produce a pair of largely similar, parallel, and opposite outflows, so-called jets.
However, whether the jets launched from AGN are actually symmetric is currently an open question.
Alongside differences in the surrounding medium between the two regions crossed by the outflows, asymmetric jet production was theoretically shown to be possible as a consequence of persistent asymmetries in the accretion disk \citep[see, for example,][]{Wang1992, Fendt2013, Nathanail2020}.
Recently, new works have shown that even in symmetrically launched jets, asymmetries in the observed physical parameters can arise from the passage of jet instabilities, such as shocks, downstream of both jets as a consequence of the different viewing angles between the two sides \citep{Fromm2018, Saiz2025}.
From an observational point of view, it is unknown whether the triggered asymmetries show significant time variability.

The magnetic field content of the relativistic jets and their central regions, along with possible asymmetries in the jet production mechanisms, can be investigated through very long baseline interferometry (VLBI) observations of jets on sub-parsec to parsec scales. For example, VLBI images may highlight differences in the velocity and collimation profiles, as well as in the overall morphology of the two jets \cite[][]{Baczko2019, Saiz2025}.
The best sources to explore the rise of jet asymmetries are double-jetted radio galaxies, namely AGN seen at large viewing angles, in which, thanks to the reduced projection effects, we have the possibility to detect both jets.

In this work, we focus on NGC\,1052, a largely studied double-sided radio galaxy for which an asymmetric jet production was previously suggested \citep{Baczko2019}.
NGC\,1052 is a nearby source, with a redshift of $z = 0.005$ \citep[][]{Koss2022} and an estimated central black hole (BH) mass of $\mathrm{M_{BH}} = 10^{8.2} \, \mathrm{M_\odot}$ \citep{Woo2002}.
The source has been largely studied in the literature by means of VLBI observations, allowing us to gain a deep insight into the properties of its jets, such as their collimation\footnote{The collimation distance is signaled by the jet transitioning from a quasi-parabolic geometry ($d = \mathrm{z}^\psi$, $\psi \sim 0.50$ with $d$ being the jet width and $\mathrm{z}$ the distance from the core) to a conical one ($\psi \sim 1$).}, propagation speed and magnetic field content \citep{Kameno2001,Vermeulen2003,Kadler2004b, Baczko2016, Baczko2019, Baczko2022, Baczko2024}.
The accretion disk in NGC\,1052 is proposed to be an advection-dominated accretion flow \citep[ADAF, see, for example,][]{Falocco2020}.
At larger scales, the nuclear region of NGC\,1052 is embedded in a geometrically thick, clumpy molecular torus with a radius of $2.4 \pm 1.3 \, \mathrm{pc}$ \citep{Kameno2020}.
The column density of the neutral hydrogen (HI) has been constrained to be between $10^{22} \, \mathrm{cm^{-2}}$ - $10^{24} \, \mathrm{cm^{-2}}$ \citep{Weaver1999, Guainazzi1999, Kadler2004a}, while more recent measurements constrain higher values for the molecular hydrogen (H2) of $(3.3 \pm 0.7) \times 10^{25} \, \mathrm{cm^{-2}}$ \citep{Kameno2020}.
The torus presents an ionized component which extends 0.1 pc ($\sim 1 \, \mathrm{mas}$) towards the eastern jet and 0.7 pc ($\sim 7.5 \, \mathrm{mas}$) towards the western one \citep{Kameno2001}. 
The latter is argued to cause free--free absorption of the jet-emitted radiation, largely seen from VLBI observations at relatively low frequencies \citep[see, for example,][]{Kadler2004b}, with previous studies inferring a free--free absorption coefficient of $\tau_\mathrm{1 GHz} \sim 300$--1000 \citep{Kameno2001, Sawada-Satoh2008} and $\tau_\mathrm{8 GHz} \sim 3$ \citep{Kadler2004b}.

In this work, we expand the previously published VLBI results on NGC\,1052 by adding new observations at 43\,GHz and 86\,GHz with the goal of exploring the persistence of asymmetries between the two jets together with further insights on their magnetic fields.
The paper is divided as follows. In Sect.\,\ref{sec:data_sec} we present our data; in Sect.\,\ref{sec:results} we present our results, focusing on the jet collimation and velocity profiles together with the (43-86)\,GHz spectral index map; in Sect.\,\ref{sec:discussion} we discuss the results, with a focus on the time variability of the jet asymmetry and magnetic field strength of the jets; in Sect.\,\ref{sec:Conclusions} we present our conclusions.
We assume a $\Lambda$CDM cosmology with $H_0 = 71 \ \mathrm{h \ km \ s^{-1} \ Mpc^{-1}}$, $\Omega_M = 0.27$, $\Omega_\Lambda = 0.73$ \citep{Komatsu}.
The luminosity distance of NGC~1052 is $D_L = 19.23 \, \mathrm{Mpc}$ \citep{Tully2013}, and 1 mas corresponds to 0.093 pc and 6134\,$R_\mathrm{S}$\footnote{$R_\mathrm{S} = 2 \mathrm{G M_{BH}} / c^2$ is the Schwarschild radius.}.
As viewing angle for NGC\,1052 we assume $\theta > 80\degree$, according to \citet{Vermeulen2003, Baczko2019, Baczko2022}.

\section{Data analysis} \label{sec:data_sec}

\subsection{Original data} \label{sec:data}

\begin{table*}[htpb]
\caption{Summary of the new VLBI images presented. The upper three lines report the properties of the new 43 GHz data, while the bottom one reports the new 86\,GHz observation.}
\centering
\begin{tabular}{llllll}
\hline
\multicolumn{1}{c}{\begin{tabular}[c]{@{}c@{}}$\nu$\\ {[}GHz{]}\end{tabular}} & 
\multicolumn{1}{c}{Obs date} & 
\multicolumn{1}{c}{\begin{tabular}[c]{@{}c@{}}Beam\\ {[}mas, mas, deg{]}\end{tabular}} & \multicolumn{1}{c}{\begin{tabular}[c]{@{}c@{}}Total flux\\ {[}Jy{]}\end{tabular}} & \multicolumn{1}{c}{\begin{tabular}[c]{@{}c@{}}Brightness peak \\ {[}Jy/beam{]}\end{tabular}} & \multicolumn{1}{c}{\begin{tabular}[c]{@{}c@{}}$\sigma$ \\ {[}mJy/beam{]}\end{tabular}} \\ 

\hline 
\hline

\multicolumn{1}{c}{43.2}  & \multicolumn{1}{c}{23 Apr 2021}  & \multicolumn{1}{c}{0.562$\times$0.23, $-$7.13} & \multicolumn{1}{c}{0.760} & \multicolumn{1}{c}{0.106} & \multicolumn{1}{c}{0.230} \\ 

\multicolumn{1}{c}{43.2}  & \multicolumn{1}{c}{01 Oct 2021}  & \multicolumn{1}{c}{0.716$\times$0.243, $-$14.1} & \multicolumn{1}{c}{1.024} & \multicolumn{1}{c}{0.098} & \multicolumn{1}{c}{0.243} \\

\multicolumn{1}{c}{43.2}  & \multicolumn{1}{c}{03 Apr 2022}  & \multicolumn{1}{c}{0.677$\times$0.235, $-$11.8} & \multicolumn{1}{c}{0.734} & \multicolumn{1}{c}{0.081} & \multicolumn{1}{c}{0.217} \\ 

\hline


\multicolumn{1}{c}{86.2}  & \multicolumn{1}{c}{03 Apr 2022}  & \multicolumn{1}{c}{0.291$\times$0.067, $-$9.26} & \multicolumn{1}{c}{0.745} & \multicolumn{1}{c}{0.620} & \multicolumn{1}{c}{0.648} \\ 

\hline

\end{tabular}
\begin{flushleft} 
\textbf{Notes.} Column 1: observing frequency in GHz; Column 2: date of the observation; Column 3: beam size and position angle; Column 4: total flux density in Jy; Column 5: brightness peak value in Jy/beam; Column 6: thermal noise in mJy/beam.
\end{flushleft}
\label{table:original_maps}
\end{table*}

In this paper, we present new observations of NGC\,1052 performed with the global millimeter-VLBI array (GMVA) \citep{Ros2024} with a six-month cadence in the period April 2021 -- April 2022 at a frequency of 86\,GHz.
The observations were accompanied by intervaled observations at 43\,GHz performed by the very long baseline array (VLBA).
Unfortunately, we are able to only present one image at 86\,GHz, namely the one performed in April 2022, given that a large number of problems afflicted the April 2021 and October 2021 epochs.
For a detail description of the problems afflicting the 86\,GHz observations alongside the amplitude scaling problems at 43\,GHz, we refer to Appendix \ref{app:problems}.

The data were calibrated using the common astronomy software application (CASA)-based pipeline rPicard \citep{Janssen2019}.
In rPicard, several parameters must be specified a priori, such as the signal-to-noise ratio thresholds for the various fringe-fitting stages, as well as the choice of reference antennas.
Consequently, to achieve an optimal calibration following this procedure, multiple runs on each dataset were performed with different combinations of the prior parameters.
In the final step, the imaging and self-calibration steps on the final calibrated dataset were performed using the standard procedure for VLBI data in difmap \citep{Shepherd1994}.

\begin{figure}[t]
    \centering
    \includegraphics[width=\linewidth]{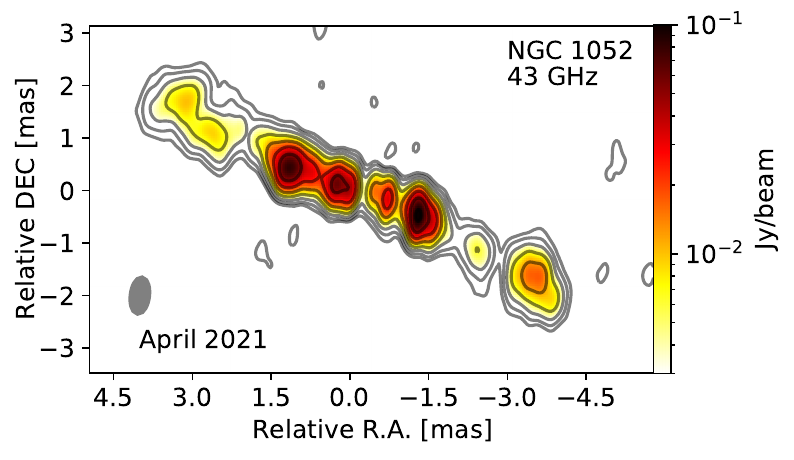}\par 
    \includegraphics[width=\linewidth]{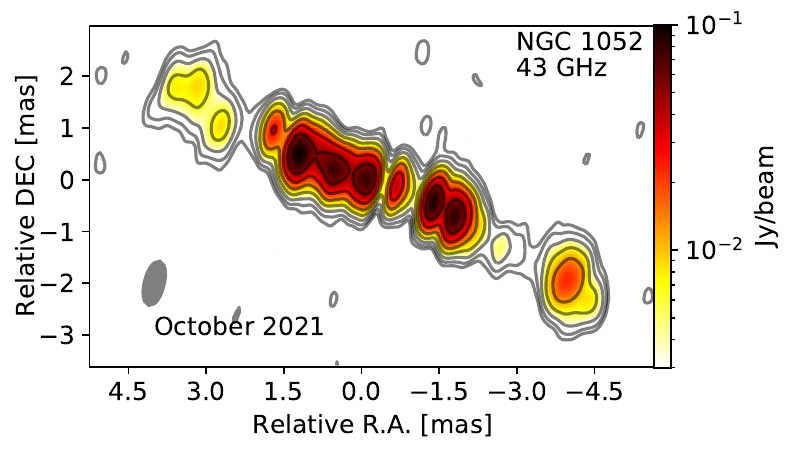}\par
    \includegraphics[width=\linewidth]{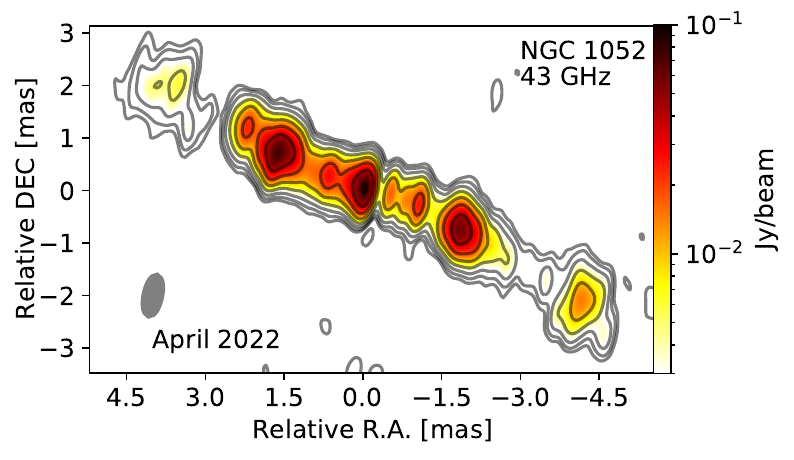}\par
    \caption{43\,GHz VLBA observations of NGC\,1052 performed in: i) April 2021 (top panel); ii) October 2021 (middle panel); April 2022 (lower panel). The contours start at 3$\sigma$, which value is reported in Table\,\ref{table:original_maps} for each map. The grey ellipse represents the naturally-weighted beam.}
    \label{fig:43GHz_maps}
\end{figure}

For the 43\,GHz data, due to their relatively high quality, only two runs were necessary in rPicard to achieve science-ready calibrated data; the resultant images are shown in Fig\,\ref{fig:43GHz_maps}.

\begin{figure*}[t]
    \centering
    \begin{multicols}{2}
    \includegraphics[width=\linewidth]{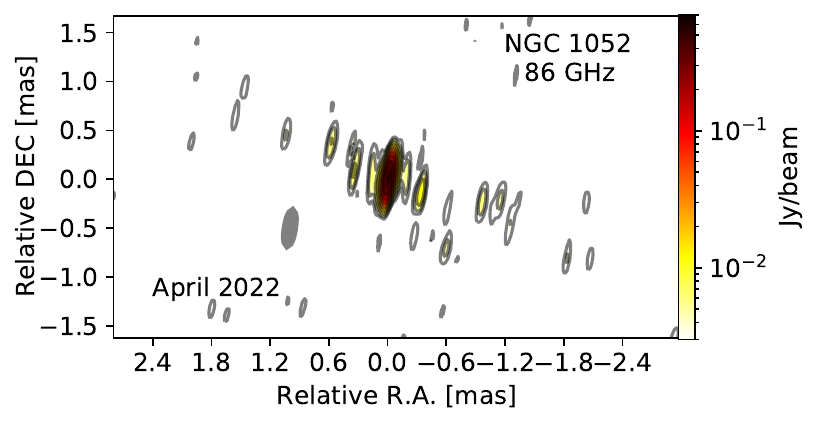}\par
    \includegraphics[width=\linewidth]{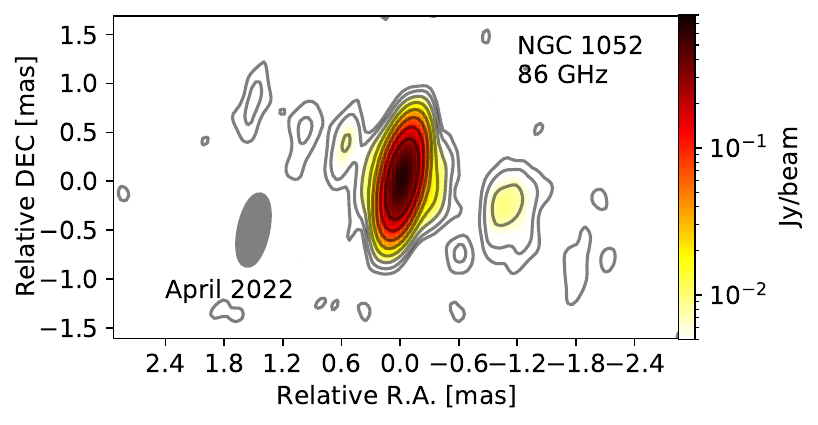}\par
    \end{multicols}
    \caption{April 2022 86\,GHz GMVA image. Left panel: uniform weighting. Right panel: image convolved with the 43\,GHz April 2022 beam ($0.677 \, \mathrm{mas} \times 0.235 \, \mathrm{mas}, -11.8\degree$). The contours start at 3$\sigma$.}
    \label{fig:86GHzmap}
\end{figure*}

The situation at 86\,GHz was more complicated, and epoch April 2022 required five runs to find a set of parameters that led to a properly calibrated dataset.
The final image is presented in Fig.\,\ref{fig:86GHzmap}: the left panel shows the uniformly weighted image, while the right panel displays the same data convolved with the 43\,GHz - April 2022 beam.
As mentioned earlier, the low data quality for epochs April and October 2021 prevented the generation of a properly calibrated data set.
In the final runs to try to recover as many baselines as possible, the SNR threshold for the science fringe fitting was lowered to 2.9 and, despite that, many antennas had a flag rate of 100\%. 

Table \ref{table:original_maps} summarizes the important parameters of the final maps used for the analysis. 
In Sect.\,\ref{sec:86GHz} we discuss extensively the quality and validity of our new 86\,GHz image.

\subsection{Gaussian modelfit components} \label{sec:gaussian}

To compute the jet kinematics (see Sect.\,\ref{sec:modelfit_kinematics}), we fit the visibilities with circular Gaussian components following the \textit{modelfit} task within difmap.
The obtained components are reported in Appendix \ref{app:model}.
As uncertainties, we assume conservative values of $10\%$ error on the flux densities and a $25\%$ error on the size.
The uncertainty on the radial distance for each component is calculated by summing in quadrature 25\% of their size together with 25\% of the circular equivalent beam.
To determine whether a certain component is below the resolution limit, and so has to be neglected, we use \citep{Kovalev2005}:
\begin{equation}
    \theta_\mathrm{lim} = b_\Theta \sqrt{\frac{2\, \mathrm{ln}\,2}{\pi} \, \mathrm{ln} \, \Bigg( \,  \frac{\mathrm{SNR}}{\mathrm{SNR} - 1} \, \Bigg)}
    \label{eq:res_limit}
\end{equation}
in which $b_\Theta$ is the full-width at half maximum (FWHM) of the beam along a certain direction $\Theta$ and $\mathrm{SNR}$ is the signal to noise ratio of each component, calculated as the component flux over the thermal noise in a nearby region which does not include jet emission.
The FWHM of the beam is assumed to be the beam size along the position angle of both jets. 

As discussed in the previous section, the 86\,GHz 2021 observations were not imaged due to their low data quality. 
However, we performed a modelfit on the limited available visibilities using a single Gaussian component to model the core. 
The resultant core component parameters are listed in Appendix \ref{app:model}, and they are found to be above the resolution limit, according to Eq.~\ref{eq:res_limit}.

\subsection{Map alignment} \label{sec:jetorigin}

When comparing jet emission at different frequencies, it is crucial to consider the so-called core-shift effect. 
Due to the opacity of the nuclear region, the VLBI core, defined as the optically thick region in the jet, moves downstream of the jet with decreasing observing frequency \citep{Blandford1979}.
Therefore, to perform a multi-frequency analysis, the radio images need to be aligned to a common origin, which is ideally the position of the central black hole.

\begin{figure}[t]
    \centering
    \includegraphics[width=0.9\linewidth]{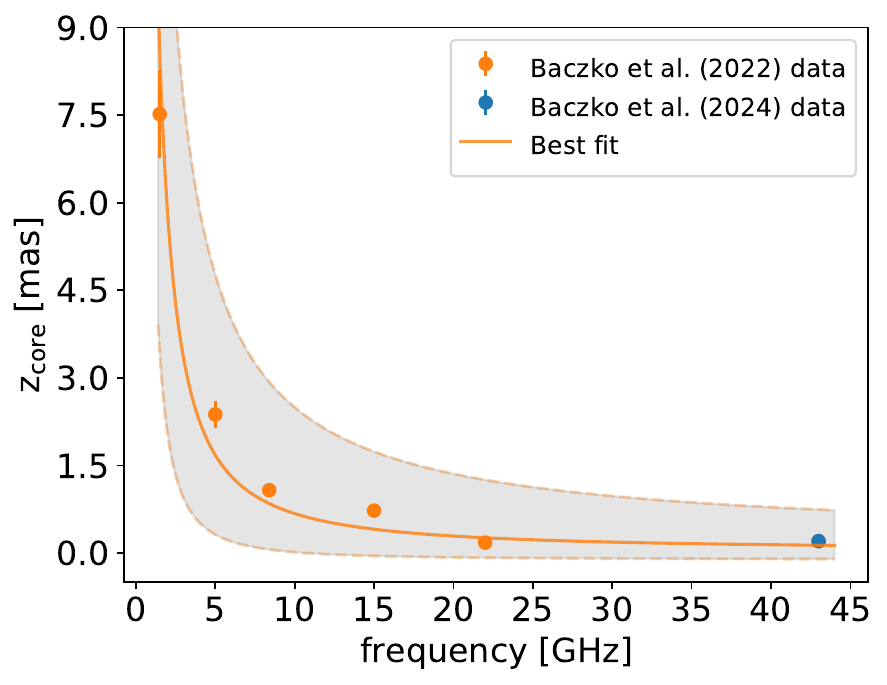}\par
    
    \caption{Core shift of the eastern jet as a function of the frequency. The orange data points are from \citet{Baczko2022} while the 43\,GHz blue data point from \citet{Baczko2024}. The orange continuous line represents the best fit, while the dashed ones represent the corresponding upper and lower limits. The grey area shows the possible best-fit line within $1\sigma$.}
    \label{fig:core_shift}
\end{figure}

The core displacement of the eastern and approaching jet between 1 GHz and 43\,GHz has been evaluated in \citet[][see their Table 3]{Baczko2022}, while the shift between the 43\,GHz and 86\,GHz cores is taken from \citet{Baczko2024}.
Recent studies \citep[for example,][]{Plavin2019} suggest the displacement to be time variable, a consequence of the varying underlying jet physical properties.
Therefore, in principle, simultaneous data are needed to adequately describe the core shift across the different frequencies.
As a consequence, in Sect.\,\ref{sec:spectralindex} we infer the 43\,GHz core shift from the new observations presented here, and we discuss the impact of the core-shift variability.
In Fig.\,\ref{fig:core_shift}, we plot the core sizes of NGC\,1052 as a function of the frequency and we perform a fit using the equation $\mathrm{z}_\mathrm{core} = c_s + a \cdot \nu^{-1/k_r}$, in which $c_s$ is the core-shift value at infinite frequency, $a$ the scaling factor, and $k_r$ the inverse of the power-law index.
As best-fit parameters, we obtain $c_s = 0.05 \pm 0.16 \, \mathrm{mas}$, $a = 14.94 \pm 7.67$, and $k_r = 0.73 \pm 0.38$.
The parameter $k_r$ is of high importance since it reflects the physical properties of the underlying jet.
In detail, in the case of a conically expanding jet in equipartition between particle and magnetic field energies, its value is expected to be $\sim 1$ \citep{Blandford1979}.
On the contrary, $k_r < 1$ is expected for a non-conical and/or accelerating jet \citep[see, for example,][]{Kravchencko2025}.

The obtained value of $k_r = 0.73 \pm 0.38$ is consistent with the theoretical expectations (although consistent also with one within 1$\sigma$). 
Indeed, only the data point at 1.5\,GHz lays in the conical jet \citep[according to][]{Baczko2024}, while the other core shift measurements are within the collimating jet region.
We performed a power-law fit also excluding the point at 1.5\,GHz, in order to consider only the purely collimating jet region, obtaining $k_r = 0.46 \pm 1.00$. 
While the best-fit value is much lower, the high uncertainty makes it comparable with the estimation obtained using all the data points.

At frequencies $\nu \lesssim 43\,\mathrm{GHz}$, namely, when the core region is heavily affected by the torus opacity, \citet{Baczko2022} attributes the first component of the western jet as the core, being the dynamical center.
We make the same choice in this work, as shown in Appendix \ref{app:model}.
On the contrary, at 86\,GHz, where the obscuration effects of the torus are expected to be minimal - if not null - the core is associated with the brightest component, which lies on the position of the brightest pixel in our model fitting.
Throughout the analysis, the images are centered on the respective core components.
In Sect.\,\ref{sec:spectralindex} we discuss whether the core shift between the 43\,GHz and 86\,GHz cores obtained in \citet{Baczko2024} is also valid for the April 2022 data presented here.

\section{Results} \label{sec:results}

\subsection{86\,GHz observations} \label{sec:86GHz}

Due to the relatively low surface brightness of the extended emission, imaging the jet in NGC\,1052 is a non-trivial task.
The interpretation of the visibilities, and thus the resulting image, is not straightforward and may be influenced by human biases during the cleaning process.
To mitigate this issue, multiple imaging attempts with slightly different approaches in difmap were carried out to obtain the final image shown in Fig.~\ref{fig:86GHzmap}.
The selected image is the one with the lowest $\sigma_\mathrm{rms}$ across all the imaging attempts and with a jet structure compatible with the 43\,GHz simultaneous observations.
In Appendix \ref{app:other86}, we show an alternative version of the imaging at 86\,GHz to prove that, even by accounting for some differences in the final reconstructed jet emission, the physical discussion and conclusions we extrapolate from the 86\,GHz data remain unaltered.



The image chosen for our analysis has a total flux of $\sim 0.745\,\mathrm{Jy}$ with a brightness peak of $0.62\,\mathrm{Jy/beam}$, largely in agreement with the properties of the 86\,GHz image shown in \citet{Baczko2024} from 2017 observations.
The core flux density is $\sim 0.650\,\mathrm{Jy}$ corresponding to $\sim 87$\% of the total flux detected.
From the ALMA monitoring, in April 2022 NGC\,1052 showed a flux of $0.994\,\mathrm{Jy}$ at $91.5\,\mathrm{GHz}$\footnote{\url{https://almascience.eso.org/sc/}}, $\sim$33\% higher than our obtained VLBI flux.
We expect most of this missing flux to come from the extended jet since the lack of short baselines of the GMVA makes it less sensitive to extended jet emission.
When performing the spectral analysis (see next section), we consider and discuss the possible implications of the missing flux.

The situation differs for the 2021 epochs.
As reported in Appendix \ref{app:model}, the April and October observations have central Gaussian components with fluxes of $1.12 \, \mathrm{Jy}$ and $1.81 \, \mathrm{Jy}$, respectively.
From the ALMA monitoring, NGC\,1052 was observed to have a total flux of $1.43 \, \mathrm{Jy}$ on 28.03.2021 and $1.214 \, \mathrm{Jy}$ on 23.10.2021.
In the first case, the flux of the core component corresponds to $\sim 80\%$ of the source flux, leading to a core compactness in agreement with the 86\,GHz observation presented here and in \citet{Baczko2024}.
In the second one, the flux is higher than the actual flux density observed, a sign of improper amplitude calibration.
This is not surprising, considering the numerous problems which afflicted the October 2021 session (see Sect.\,\ref{sec:data}).

We calculate the brightness temperature of the 86\,GHz core components using Eq.\,5 in \citet{Kadler2004b}.
For the October 2021 epoch, we use $80\%$ of the ALMA monitoring flux as the component flux, consistent with the compactness of the April 2021 observation.
We obtained brightness temperatures of $T_\mathrm{b} = (1.25 \pm 0.51) \times 10^{11} \, \mathrm{K}$, $T_\mathrm{b} = (1.30 \pm 0.54) \times 10^{11} \, \mathrm{K}$, and $T_\mathrm{b} = (1.30 \pm 0.47) \times 10^{13} \, \mathrm{K}$ for the April 2021, October 2021, and April 2022 epochs, respectively.
Brightness temperatures of $\sim 10^{11} \, \mathrm{K}$ at 86\,GHz are consistent with the values reported at the same frequency in \citet{Baczko2024} and are larger than the value at 230\,GHz ($T_\mathrm{b} = 4.7 \times 10^9 \, \mathrm{K}$).
The exceptionally large $T_\mathrm{b}$ obtained in April 2022, which exceeds the inverse Compton limit, is most likely unrealistic.
This large value is due to the very small core component, which is most probably caused by improper sampling of the nuclear region.

\subsection{Jet structure} \label{sec:jet_structure}

\begin{figure*}[t]
    \centering
    \includegraphics[width=\linewidth]{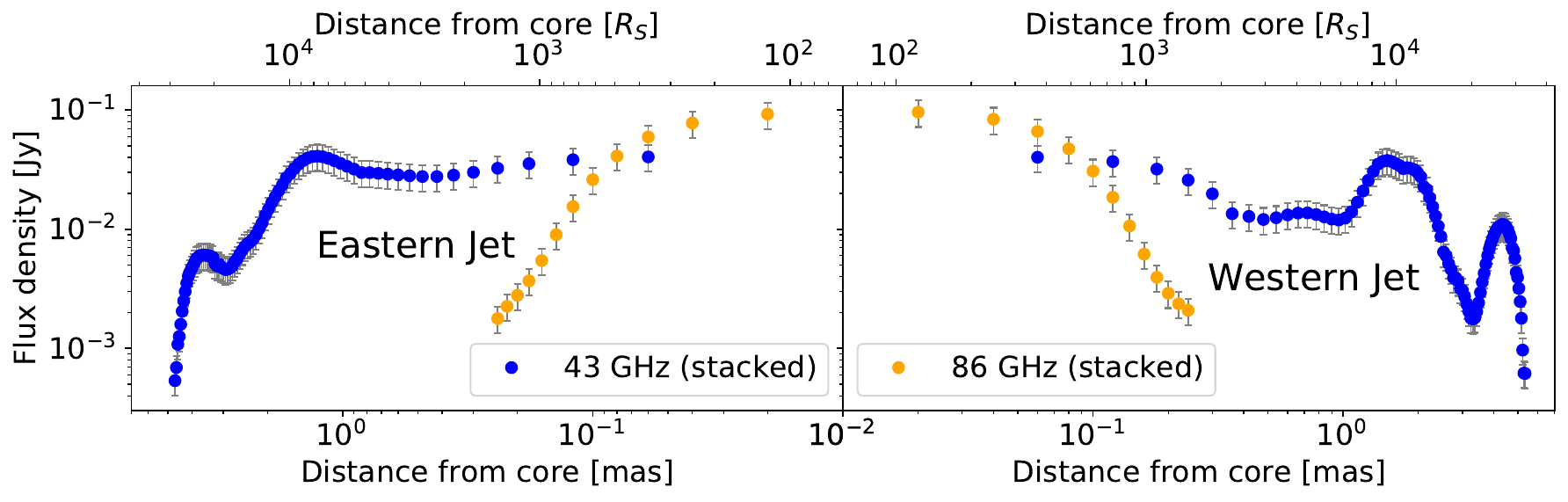}\par
    \caption{Flux density profile for the 43\,GHz and 86\,GHz stacked images for both the eastern and western jet. The double-sided jets show a high level of symmetry at 86\,GHz, while a dimmer western jet on sub-mas scales is visible at 43\,GHz.}
    \label{fig:fluxprofiles}
\end{figure*}

The 43\,GHz images of NGC\,1052 shown in Fig.\,\ref{fig:43GHz_maps} lack a central, bright core component at the location of the jet nozzle, which, in contrast, is clearly visible in the 86\,GHz image (see Fig.\,\ref{fig:86GHzmap}).
This feature, which is unusual for AGN jets observed with VLBI, is consistent with previous 43\,GHz observations reported by \citet{Baczko2022}. 
Specifically, the source exhibited a prominent central component at 43\,GHz in data obtained between 2005 and 2007, whereas more recent images from 2008–2009 show a progressive disappearance of this bright core.
The absence of a bright, central component at 43\,GHz in certain epochs is reproduced in the relativistic hydrodynamic simulations of an NGC\,1052-like source presented in \citet{Saiz2025}. 
In their work, the phenomenon is attributed to free--free absorption caused by the torus and the injection of traveling shocks downstream the jets. 
Additionally, all three 43\,GHz images show a bridge of fainter brightness and apparent thinner width between the eastern jet core and the western jet.
This structure is well known at lower frequencies \citep[see, for example,][]{Kadler2004b} to be a consequence of the impact of the torus, which therefore remains the main candidate also at 43\,GHz.
Further evidence in this direction is given by the presence of free--free absorption caused by the ionized torus in the 43\,GHz images, as well as its impact on the observed jet morphology. 
We discuss this in Sects.\,\ref{sec:spectralindex} and \ref{sec:jetmorph}.

As seen in Fig.\,\ref{fig:43GHz_maps}, the launched jets show some degree of asymmetries, with high-intensity regions whose positions change across the epochs.
Focusing on the eastern jet, in April 2021, the jet shows two bright spots connected by a relatively faint bridge of radio emission.
In the consecutive epoch, October 2021, this bridge greatly increased its luminosity, matching that of the two neighboring regions.
Finally, in April 2022, we recovered a morphology similar to that observed one year before, with the first bright spot closer to the core remaining relatively steady while the second moved downstream the jet at a distance of $\sim 1.8\,\mathrm{mas}$ in R.A.
The western jet also shows a high level of variability, with two bright regions within 2 mas from the central engine in the October 2021 epoch and only one bright spot in April 2021 and 2022.
The variability described could, in principle, be associated with the time-variable torus and the traveling of jet components downstream of both jets \citep{Saiz2025}.
We analyze and discuss in detail the knots kinematics in Sect.\,\ref{sec:modelfit_kinematics}.

In Fig.\,\ref{fig:fluxprofiles} we present the flux density profiles of the stacked images at 43\,GHz and 86\,GHz, which are shown in Appendix \ref{app:stacked}.
The 43\,GHz profiles (blue data points) show asymmetries between the two jets.
Within the first mas, the western jet appears dimmer with respect to the eastern counterpart, while at a larger distance from the core the two jets switch, with the recessing one showing a region of enhanced brightness at $\sim -4 \,\mathrm{mas}$. 
In Sect.\,\ref{sec:jetmorph} we discuss whether this asymmetry is a consequence of intrinsic differences in the jet launching.
On the contrary, the 86\,GHz observations reveal a different scenario with the flux density profiles (orange data points) showing an high level of symmetry within $\sim 0.3 \, \mathrm{mas}$.


\subsubsection{Collimation profile} \label{sec:collimation}

\begin{figure*}[t]
    \centering
    \includegraphics[width=\linewidth]{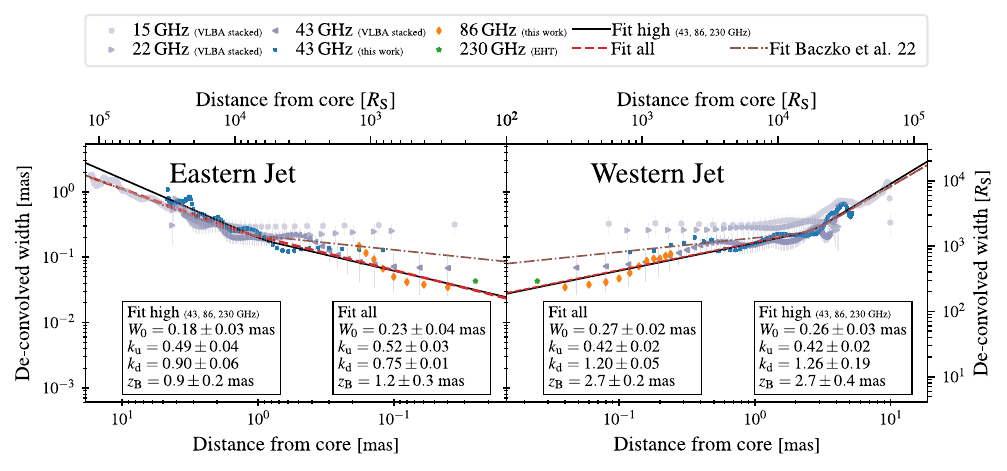}\par
    \caption{De-convolved jet width as a function of the distance from the core for the eastern and western jet. This plot, originally published in \citet{Baczko2024}, is updated with the new profiles at 43\,GHz and 86\,GHz extrapolated in this work.
    We show three different fits: i) black continuous line, best-fit line for the high frequency data (43, 86, and 230\,GHz); ii) red dashed line, best-fit line for all the data; iii) brown dot-dashed line, best-fit line reported in \citet{Baczko2022}. In the tables, $W_0$ are the initial jet widths, $k_\mathrm{u}$ and $k_\mathrm{d}$ are the upstream and downstream power law indexes, respectively, and $\mathrm{z}_\mathrm{b}$ are the transition distances.}
    \label{fig:coll}
\end{figure*}

In Fig.\,\ref{fig:coll}, we present the collimation profile shown in \cite{Baczko2024} updated with the 43\,GHz and 86\,GHz stacked jet width.
The new data points have been extrapolated using the pixel-by-pixel method presented in \cite{Ricci2022}.
For the details on the code, we refer to Sect.~2 in the paper. 
Here we highlight how the code allows us to extrapolate the integrated flux, jet width, and opening angle for each consecutive jet slice perpendicular to the jet direction both in the jet (the eastern one) and in the counter-jet (the western one).
The de-projected width is computed as $w = \sqrt{w_m^2 - \theta^2}$ in which $w_m$ is the width of the fitted Gaussian profile and $\theta$ is the FWHM of the beam.
For the uncertainties on the jet width, we assume a conservative value of $20\%$ of the extrapolated jet width, which accounts also for the fitting errors, summed in quadrature with $10\%$ of the equivalent circular beam.

As seen in Fig.\,\ref{fig:coll}, the new profiles coincide with the previous estimations of the jet width (shown as gray points).
On the one hand, the new 43\,GHz jet widths are largely in agreement with the previous data points at the same frequency and match within $1\,\sigma$.
On the other hand, the 86\,GHz data points show a slightly thinner jet upstream $\sim 0.1 \, \mathrm{mas}$ than the one recovered at 43\,GHz (although consistent within $1\,\sigma$ with the previously obtained 43\,GHz data points) and are in agreement with the width recovered from the EHT observation at 230\,GHz \citep{Baczko2024}. 
Downstream $\sim 0.1 \, \mathrm{mas}$ the 86\,GHz data points highly match the ones recovered at 43\,GHz.
We notice how the jet widths from the modelfit of the core components of the 2021 86\,GHz observations are consistent with the measurement obtained from the stacked image at the same frequency (see Appendix \ref{app:stacked}).
On sub-mas scales, the jet becomes progressively thinner at increasing observing frequencies, which we suggest to be a consequence of a progressively less important scattering caused by the torus. 
Indeed, the wider jet at relatively low frequencies ($\nu \leq 22\,\mathrm{GHz}$) could be a consequence of the emitted jet radiation being highly scattered by the torus, which leads, in turn, to the cylindrical geometry observed and the deviation from the parabolic expansion, as previously hypothesized in \citet{Baczko2022}.
On the contrary, at higher frequencies ($\nu \geq 43\,\mathrm{GHz}$), the scattering caused by the torus becomes progressively negligible and the true jet width profile is recovered.
Overall, the newly added data points confirm the scenario of the jet expanding initially with a parabolic regime.

We fit a broken power-law to both jets described by:
    \begin{equation}
     d(\mathrm{z}) = W_0 2^{(k_\mathrm{u}-k_\mathrm{d})/s}\left(\frac{\mathrm{z}}{\mathrm{z}_\mathrm{b}}\right)^{k_\mathrm{u}}\left[1+\left(\frac{\mathrm{z}}{\mathrm{z}_\mathrm{b}}\right)^s\right]^{(k_\mathrm{d}-k_\mathrm{u})/s} \, ,
    \end{equation}
in which $W_0$ is the initial jet width, $k_\mathrm{u}$ and $k_\mathrm{d}$ are the upstream and downstream power law indexes, respectively, $s$ is the hardness of the broken profile, and $\mathrm{z}_\mathrm{b}$ is the distance at which the break occurs.
At first, we fit all the data available for both jets (see Fig.\,\ref{fig:coll}).
For the eastern jet, we derive an initial jet width of $W_0 = 0.23 \pm 0.04 \, \mathrm{mas}$, with upstream and downstream power-law indexes of $k_\mathrm{u} = 0.52 \pm 0.03$ and $k_\mathrm{d} = 0.75 \pm 0.01$, respectively.
For the western jet, the corresponding values are $W_0 = 0.27 \pm 0.02 \, \mathrm{mas}$ $k_\mathrm{u} = 0.42 \pm 0.02$ and $k_\mathrm{d} = 1.20 \pm 0.05$.
The transition from a parabolic to a conical jet shape is constrained to occur at $\mathrm{z_b} = 1.2 \pm 0.3 \, \mathrm{mas}$ for the eastern jet and $\mathrm{z_b} = 2.7 \pm 0.2 \, \mathrm{mas}$ for the western one.
The inclusion of new data points in this work allows for more accurate sampling of the sub-mas scales, revealing that the jet geometry steepens closer to the core more than previously constrained.
This finding aligns with the results suggested by \citet{Baczko2024}, aided by the newly added EHT measurement at 230\,GHz.
On the contrary, in \citet{Baczko2022}, upstream indexes of $k_\mathrm{u} = 0.22 \pm 0.06$ and $k_\mathrm{u} = 0.21 \pm 0.05$ were derived for the approaching and receding jets, respectively (shown as brown dashed dotted lines in Fig.\,\ref{fig:coll}).
For both the eastern and western jets, both the transition distance and the conical jet geometry obtained here are largely in agreement with those of \citet{Baczko2022}.
Additionally, we notice that along the western jet, we detected a longer transition distance which leads to a jet width expansion rate steeper than the one inferred for the eastern jet.
Interestingly, the theoretical work presented in \citet{Saiz2025} shows a very similar behavior: a jet break occurring farther from the core results in a larger downstream expansion rate.
We also performed a separate fit using only the high-frequency data presented here (43\,GHz and 86\,GHz) together with the 230\,GHz point (black continuous line in Fig.\,\ref{fig:coll}), which yields consistent results with the previous fit.
The best-fit parameters for the different scenarios are shown in Appendix \ref{app:model}.

Overall, we propose the fit to all data points (dashed red line in Fig.\,\ref{fig:coll}) as the best representation for both jet geometries. 
For the eastern jet, it is the only fit that adequately captures the parabolic expansion on sub-mas scales and the quasi-conical expansion downstream of the transition distance, while for the western, it largely matches with the fit proposed in \citep{Baczko2022} while describing the parabolic region in a more accurate way.
Accordingly, throughout the rest of the paper, we will adopt the following parameters.
For the eastern jet: $k_u = 0.52 \pm 0.03$, $k_d = 0.75 \pm 0.01$, and $\mathrm{z_b} = 1.2 \pm 0.3 \, \mathrm{mas}$; while for the western one: $k_u = 0.42 \pm 0.02$, $k_d = 1.20 \pm 0.05$, and $\mathrm{z_b} = 2.7 \pm 0.2 \, \mathrm{mas}$.
The expansion rates we extrapolate are in agreement with the theoretical predictions from \citet{Saiz2025}. Overall, our findings indicate that on sub-mas scales the jets appear symmetric, showing similar expansion rates and initial jet widths, while asymmetries arise further downstream, with the western jet showing a larger transition distance and a much steeper conical profile than the eastern one.

\subsubsection{Jet speed} \label{sec:modelfit_kinematics}

\begin{figure*}[t]
    \centering
    \begin{multicols}{2}
    \includegraphics[width=0.9\linewidth]{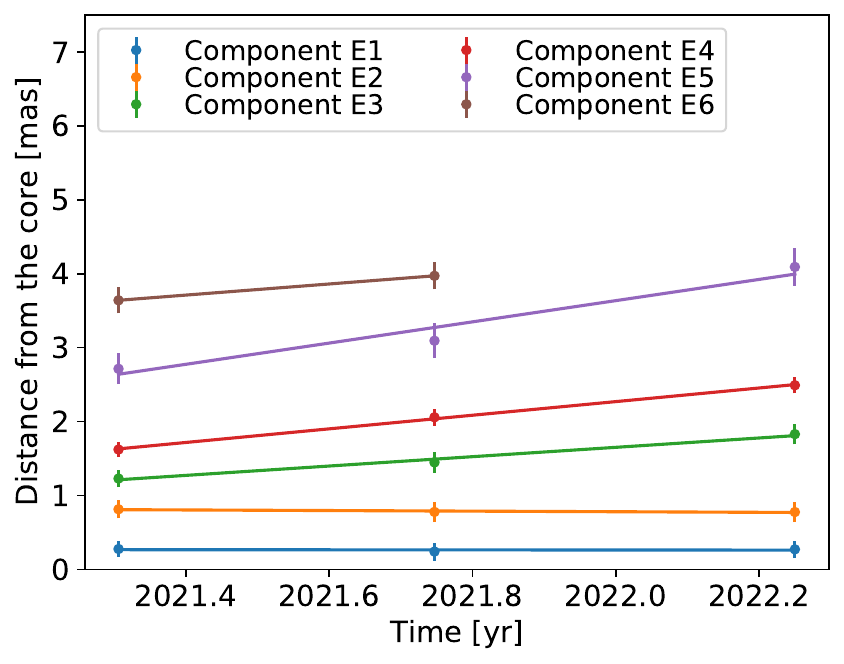}\par
    \includegraphics[width=0.9\linewidth]{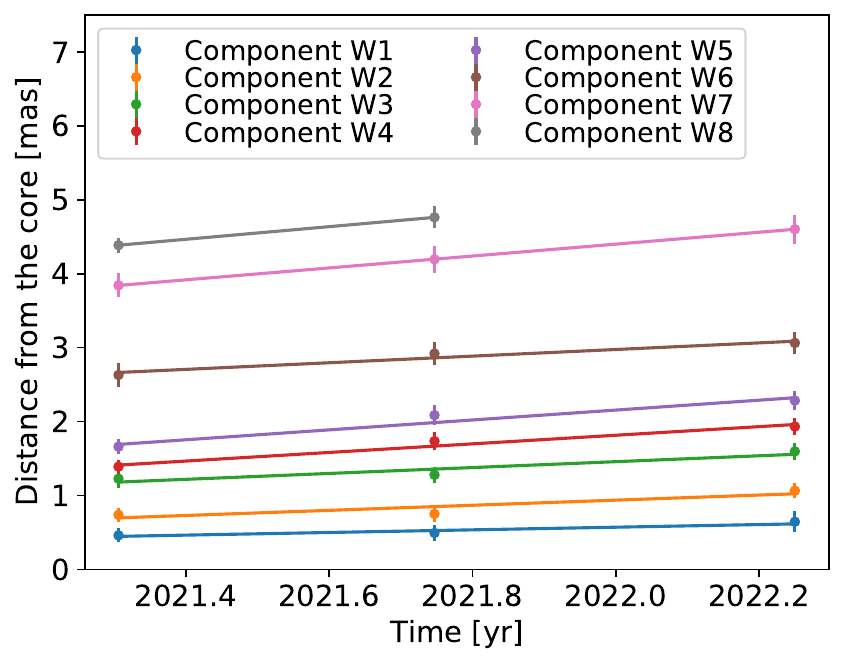}\par
    \end{multicols}
    \caption{Distance from the core across the three epochs between April 2021 and April 2022 for the Gaussian components reported in Appendix \ref{app:model}. Left panel: eastern jet. Right panel: western jet. The different colored continuous lines are the best-fit curves for the different components.}
    \label{fig:jet_speed}
\end{figure*}

To calculate the jet speed, we cross-correlate the modelfit components across the three different epochs at 43\,GHz (see Appendix \ref{app:model} for the list of the jet components).
In Fig.\,\ref{fig:jet_speed}, we present the displacement of each component for the eastern (left panel) and the western jet (right panel).
For each individual component, we perform a linear fit of the form $\mathrm{z}_\mathrm{c} = a + \mu \, \mathrm{t}$, where $a$ is the initial position, $\mathrm{z}_\mathrm{c}$ is the distance from the core, $\mu$ is the jet speed in units of mas/yr and $\mathrm{t}$ is the time in years.
The best-fit value of $\mu$ for each component, expressed as a fraction of the speed of light, is reported in Table \ref{tab:jet_speed}.
In both jets, we detect the speed of the components to increase with the distance from the core, as observed for the majority of AGN jets \citep[see, for example][]{Lister2021}.
Specifically, the components reach the speed $\beta = \mu/\mathrm{c} \sim 0.25$ on both jets at $\sim 4 \, \mathrm{mas}$ from the core.
The only exception is represented by component E5, which shows an apparent speed of $\beta = 0.41 \pm 0.09$.
It is, however, possible that what we call component E5 detected in epoch 2022 is actually a combination of components E5 and E6, and it is located halfway between the two.
Indeed, as visible from Fig.\,\ref{fig:43GHz_maps}, bottom panel, the easternmost jet region does not show a clear double-component structure, visible on the contrary in the two previous epochs, likely due to a consequence of a lower SNR in the region or evolving jet morphology.
In this scenario, the higher velocity of component E5 would be a consequence of an improper cross-identification and not representative of the actual speed at such distance.
We note that the first two components in the eastern jet are not moving and can be associated with recollimation shocks, in agreement with the findings of \citet{Saiz2025}.

The jet speeds inferred in this work are smaller than the ones inferred in \citet{Baczko2019} between 2005 and 2009, in which average speeds of $0.343 \pm 0.037 \, \mathrm{c}$ and $0.529 \pm 0.038 \, \mathrm{c}$ have been inferred for the western and eastern jet, respectively.
In contrast, in this work we infer average speeds of $0.19 \pm 0.01 \, \mathrm{c}$ and $0.26 \pm 0.01 \, \mathrm{c}$, respectively.
Additionally, the previously detected asymmetry in the jet speed between the approaching and receding jet is not recovered between 2021 and 2022.
Overall, this analysis might indicate how the components in both jets are moving at lower and more symmetric speeds with respect to the past, even if the result might be biased by the lower number of epochs employed for the analysis.

\begin{table}[]
\caption{Jet speed extrapolated from the circular gaussian components in the time range April 2021 - April 2022 for the eastern and western jet.}
\begin{tabular}{cccc}
\hline
\begin{tabular}[c]{@{}c@{}}Component\\ ID - eastern\end{tabular} & \begin{tabular}[c]{@{}c@{}}$\beta_\mathrm{app}$\\ {[}c{]}\end{tabular} & \begin{tabular}[c]{@{}c@{}}Component\\ ID - western\end{tabular} & \begin{tabular}[c]{@{}c@{}}$\beta_\mathrm{app}$\\ {[}c{]}\end{tabular} \\ \hline \hline
E1                                                               & $0.0 \pm 0.0$                                                          & W1                                                               & $0.05 \pm 0.02$                                                        \\ 
E2                                                               & $0.0 \pm 0.0$                                                          & W2                                                               & $0.10 \pm 0.05$                                                        \\ 
E3                                                               & $0.19 \pm 0.02$                                                        & W3                                                               & $0.12 \pm 0.05$                                                        \\ 
E4                                                               & $0.27 \pm 0.01$                                                        & W4                                                               & $0.17 \pm 0.03$                                                        \\ 
E5                                                               & $0.41 \pm 0.09$                                                        & W5                                                               & $0.20 \pm 0.04$                                                        \\ 
E6                                                               & $0.22$                                                                 & W6                                                               & $0.14 \pm 0.03$                                                        \\ 
                                                                 &                                                                        & W7                                                               & $0.24 \pm 0.01$                                                        \\ 
                                                                 &                                                                        & W8                                                               & $0.25$                                                                 \\ \hline
\end{tabular}
\begin{flushleft} 
\textbf{Notes.} Column 1: Component ID for the eastern jet; Column 2: Apparent speed of the respective component in units of speed of light; Column 3: Component ID for the western jet; Column 4: Apparent speed of the components in units of speed of light.
\end{flushleft}
    \label{tab:jet_speed}
\end{table}

\subsection{Spectral index map} \label{sec:spectralindex}

\begin{figure}[t]
    \centering
    \includegraphics[width=\linewidth]{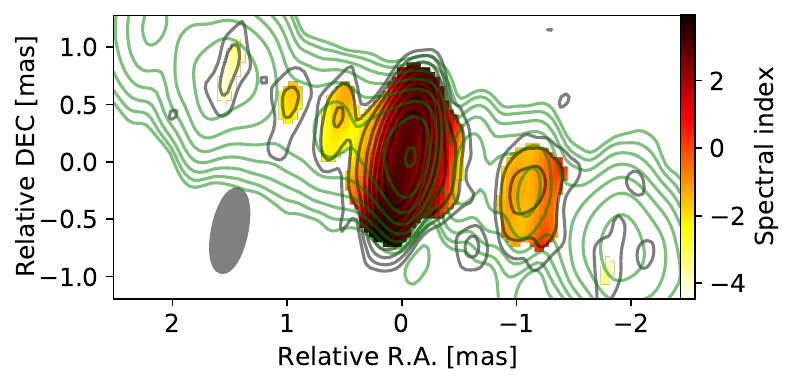}\par
    \caption{ (43--86)\,GHz spectral index maps of NGC\,1052 obtained using the core shift proposed in this work. The black contours are for the 86\,GHz emission, while the green ones are for the 43\,GHz image.}
    \label{fig:sm}
\end{figure}

In this section, we present the first (43--86)\,GHz spectral index map of NGC\,1052 (we use the formalism $S_\nu \propto \nu^\alpha$).
To obtain it, we constrained the visibilities in both datasets to the same $(u,v)$-range, we convolved the final 86 GHz image with the beam of the 43 GHz one (see Table \ref{table:original_maps}), and applied a flux cutoff at $5\sigma$. 
The pixel size was set to 0.041 mas.
To properly compute the spectral index map, the correct displacement between the 43\,GHz and 86\,GHz images in April 2022 had to be determined.
Initially, we aligned the two frequencies using the core shift obtained in \citet{Baczko2024}.
As shown in Appendix \ref{app:othersm}, the two maps appear not to be correctly aligned.
As a consequence, we attempted to align the images at the two frequencies by employing an automatic 2D cross-correlation analysis, similar to what has been done in \citet{Fromm2013, Baczko2022}.
However, obtaining meaningful results was impossible due to the poorly extended and weak jet emission recovered at 86\,GHz.
To perform the alignment we had to rely on the high-brightness region at a distance of $\sim -1\,\mathrm{mas}$ along the western jet found at both frequencies.
The shift that allows to align the contours at the two frequencies is $(x,y) = (1\pm1,0\pm1)\,\mathrm{pixels} = (0.041 \pm 0.041,0.000 \pm 0.041)\,\mathrm{mas}$; we show the correspondent spectral index map in Fig.\,\ref{fig:sm}.
The shift along the R.A. direction is positive, indicating that the 43 GHz map is shifted towards the West. 
This is unexpected, as the radio core at progressively lower frequencies is typically expected to move downstream along the approaching jet, namely the eastern one. 
However, a similar scenario was observed in \citet{Baczko2022} when calculating the alignment between 15 GHz and 22 GHz. 
A possible explanation could be that the shift, being minimal at these frequencies, could oscillate due to different conditions in the underlying jet.
As uncertainty in our alignment method, we adopt a displacement of one pixel, corresponding to the minimum calculable shift.
In Appendix \ref{app:othersm} we present alternative spectral index maps obtained by applying all possible shift combinations within a $\pm 1 \, \mathrm{pixels}$ range.
We highlight that the main physical properties discussed in the consecutive paragraphs remain unchanged across the different spectral index maps, being the values of $\alpha$ across the different regions consistent within the errors.

We find a highly inverted core region, with $\alpha_\mathrm{thick} \sim (3.0$--$3.3)$ with $\sigma = 0.16$, according to the different displacements.
The measured spectral index exceeds the theoretical limit expected from synchrotron self-absorption ($\alpha_\mathrm{thick} = 2.5$, under the assumption of the radiation emitted from a single emitting region), further suggesting that free–free absorption from the ionized torus plays a role at 43\,GHz.
This value of the spectral index is in agreement \citet{Baczko2024}, who report a spectral index of $\alpha_\mathrm{thick} = 3.27$ between 43\,GHz and 86\,GHz for the core component.
Detecting free--free absorption at such high frequencies is relatively surprising, and we further discuss it in Sect.\,\ref{sec:jetmorph}.

Along the eastern jet, the spectral index goes down to $\sim~-2$ outside of the core with the first component located at $\sim~0.5 \, \mathrm{mas}$, which shows an average spectral index of -1.9 and a gradient perpendicular to the jet direction.
The latter could be a consequence of an improper alignment for the eastern jet or the bulk of the jet emission at 86\,GHz having a slightly different position angle than the one at 43\,GHz.
We notice that with different alignments (see, for example, Fig.\,\ref{fig:othersm_align} first row, left column), such a gradient can strongly attenuate. 
The two more downstream regions inferred in the spectral index map have average index values of $-1.97 \pm 0.30$ and $- 3.89 \pm 0.29$, respectively.
Along the western jet, the spectral index goes down to $\sim -1.4$ outside of the core with the inverted spectrum region being more extended than the eastern counterpart, which we associate with the free-free absorption caused by the torus since it coincides with the bridge of absorbed emission mentioned in the previous section.
The component recovered at $\sim -1 \, \mathrm{mas}$, shows an average spectral index of $\alpha_\mathrm{thin} = -0.98 \pm 0.24$ and a gradient parallel to jet direction.
This is likely a consequence of a region in the 43\,GHz jet with reduced brightness in between two bright spots (see Fig.\,\ref{fig:43GHz_maps}, lower panel).
Concerning the very steep spectral index values obtained (down to $\sim -3.9$), while steep spectra on sub-parsec scales are recovered on similar distances in other sources \citep[down to $\sim -2.5$, see, for example,][]{Ro2023, Ricci2025}, in NGC\,1052 this may instead be a consequence of missing flux at 86\,GHz, which, as mentioned in Sect.~\ref{sec:86GHz}, is expected to originate from the extended emission.
In the scenario where a fraction of the missing flux comes from the core, the spectrum of the optically thick region would be even larger than $\alpha_\mathrm{thick} \sim (3.0$--$3.3)$, strengthening the idea of the radiation at 43\,GHz being absorbed by the ionized torus. 
Additionally, in Appendix \ref{app:othersm} we report and discuss the (43--86)\,GHz spectral index map obtained employing the alternative 86\,GHz image shown in Appendix \ref{app:other86}.

Finally, using the (43--86)\,GHz displacement of $(x,y) = (0.041 \pm 0.041,0.0 \pm 0.041)\,\mathrm{mas}$, the resulting power-law index for the core shift becomes $k_r = 0.87 \pm 0.10$, consistent with the value of $k_r = 0.73 \pm 0.38$ inferred in Sect.\,\ref{sec:jetorigin}.
We will use $k_r = 0.87 \pm 0.10$ to extrapolate the magnetic field strength in Sect.\,\ref{sec:mad}.
\section{Discussion} \label{sec:discussion}

\subsection{Constraints on the symmetric jet production} \label{sec:jetmorph}

The flux density profiles of the eastern and western jets shown in Fig.\,\ref{fig:fluxprofiles}, exhibit distinct behaviors at the two frequencies.
At 86\,GHz, the jets appear symmetric within $\sim 0.3\,\mathrm{mas}$ from the core, with fluxes consistent within uncertainties at each given distance.
At 43\,GHz, the flux density profiles differ, with the eastern jet being brighter than the western one within the inner $\sim 1 \, \mathrm{mas}$ from the core.
To investigate whether the observed asymmetry could be attributed to absorption by the ionized torus, we calculate the required optical depth for free–free absorption and the corresponding column density of the ionized material. 
These values are then compared with those reported in the literature.
We work under the assumption that the decrease in brightness in the western jet is only due to the free--free absorption from the torus, and that the jets are intrinsically symmetric (as suggested by the 86\,GHz profiles). 
Between $(0.3$--$0.8)\,\mathrm{mas}$, namely the region in which we see the reduced brightness, the average flux densities are $27\,\mathrm{mJy}$ and $12\,\mathrm{mJy}$ for the eastern and western jet, respectively.
Employing $S_\nu^{\mathrm{obs}} = S_\nu^{\mathrm{int}} \mathrm{e}^{-\tau_\mathrm{f}}$, we constrain an optical depth for the free--free absorption of $\tau_\mathrm{f} = 0.8$.
The free--free optical depth is expected to evolve with the frequency as $\propto \nu^{-2.1}$ \citep[see, for example,][]{Kadler2004b}.
In \citet{Kadler2004b} the authors inferred $\tau_\mathrm{f} \sim 3.0$ at $8.4\,\mathrm{GHz}$, which would lead to $\tau_\mathrm{f} \sim 0.09$ at $43\,\mathrm{GHz}$, one order of magnitude lower than the one we infer here.
Such a discrepancy could be attributed to the absorption being generated in two different regions at the two frequencies.
In detail, the lower frequency emission is produced in a more extended region, which could suffer absorption by a diffuse screen, whereas the absorption of the more compact area at 43\,GHz could be generated by some denser clump towards our line-of-sight, as justified later in the text.
Additionally, the value of $\tau_\mathrm{f} \sim 3.0$ at $8.4\,\mathrm{GHz}$ is in principle an average of a much larger region with respect to the one inferred at 43\,GHz, due to the reduced resolution at the lower frequency.
An alternative explanation could, in principle, involve temporal variations in the absorption caused by the torus, a plausible scenario in the case of a clumpy structure.
A clumpy torus has indeed been suggested to surround NGC\,1052 from ALMA observations \citep{Kameno2020}.

Thanks to the knowledge of $\tau_\mathrm{f}$, we can calculate the column density employing the equations shown in \citet[][]{Lobanov1998}:
\begin{equation}
    \tau_\mathrm{f} = 5 \times 10^{16} \bar{g}\, n_e\, n_i \Bigg[\frac{T}{\mathrm{K}}\Bigg]^{-1.5} \Bigg[\frac{\nu}{\mathrm{Hz}}\Bigg]^{-2} \Bigg[\frac{L}{\mathrm{pc}}\Bigg] \, \mathrm{cm^{-3}} \, ,
\end{equation}
in which $n_e = n_i$ are the numerical density of the electron and ions, which we assume to coincide, $T$ is the temperature of the torus, $L$ is the path length of the light, and $\bar{g}$ is the free--free Gauge factor.
We approximate the latter as $\bar{g} = 1$, according to Table 5 shown in \citet{van2014}.
As parameters, we use $T = 10^4\,\mathrm{K}$, $L=0.3\,\mathrm{pc}$ \citep[following][]{Kadler2004b}, $\tau_\mathrm{f} = 0.8$, and $\nu = 43\, \mathrm{GHz}$, obtaining $n_e = 3.1 \times 10^{5} \, \mathrm{cm^{-3}}$ and, consequently, a column density of $2.9 \times 10^{23} \, \mathrm{cm^{-2}}$.
However, temperatures up to $T = 10^4\,\mathrm{K}$ are expected to be large enough to sublimate the dust in the torus, going against the idea of a clumpy structure \citep[see, for example,][]{Kishimoto2007}.
By assuming a temperature one order of magnitude lower (expected to be smaller than the temperature needed for the sublimation), namely $T = 10^3\,\mathrm{K}$, we obtain a column density of $5.2 \times 10^{22} \, \mathrm{cm^{-2}}$.
Both the column densities constrained are in agreement (although not considering different possible values for the ionization and metallicity of the torus) with X-ray observations for the hydrogen column density $n_\mathrm{H} \sim 10^{22} \,\mathrm{cm^{-2}}$-$\sim 10^{24} \,\mathrm{cm^{-2}}$ \citep{Weaver1999,Guainazzi1999,Kadler2004a}, supporting the idea of absorbed emission at 43\,GHz.

Observing free--free absorption at the high frequencies of 43\,GHz is somewhat unexpected.
Nonetheless, in support of this scenario we found: i) the inferred column density is in agreement with the one independently determined with X-ray observations; 
ii) the (43-86)\,GHz spectral index map presents values of $\alpha_\mathrm{thick} > 2.5$ indicative of free--free absorption.
To explain the reduced brightness along the western jet, alternatives include Doppler de-boosting and/or an intrinsic asymmetric jet production.
On the one hand, the large viewing angle of NGC\,1052 ($\theta \geq 80\degree$) and the low velocities inferred at such distance (see Sect.\,\ref{sec:modelfit_kinematics}) cannot lead to such Doppler de-boosting.
On the other hand, the hypothesis of asymmetric jet production does not match with the symmetry between the two jets seen at 86\,GHz. 

In conclusion, we propose that the jets in NGC\,1052 are produced intrinsically with a high level of symmetry.
Asymmetries in the jet morphologies, as the one observed in Fig.\,\ref{fig:43GHz_maps} and in \citet{Baczko2019}, are suggested to be a consequence of the interplay between a clumpy, time-variable torus \citep[as shown in][]{Kameno2020}, predominantly affecting the western jet, and observational effects. 
The latter are theorized in \citet{Saiz2025}, in which the authors demonstrate that intrinsically symmetric jets can appear asymmetric at certain epochs due to the downstream traveling of perturbations injected at the jet nozzle.
These apparent asymmetries arise as a consequence of relativistic effects, such as differences in light travel time and viewing angle, which affect the apparent timing and brightness of the features in the two jets. 
\subsection{Compact scales magnetic fields} \label{sec:mad}

In this section, we use the observational parameters reported in Sect.\,\ref{sec:results} and interpret them in the context of theoretical models to investigate the magnetic field properties of the jets and assess whether the accretion disk in NGC\,1052 has reached a MAD state. 
To do this, we follow an approach similar to the one shown in \citet{Ricci2022}.

\begin{figure}[t]
    \centering
    \includegraphics[width=\linewidth]{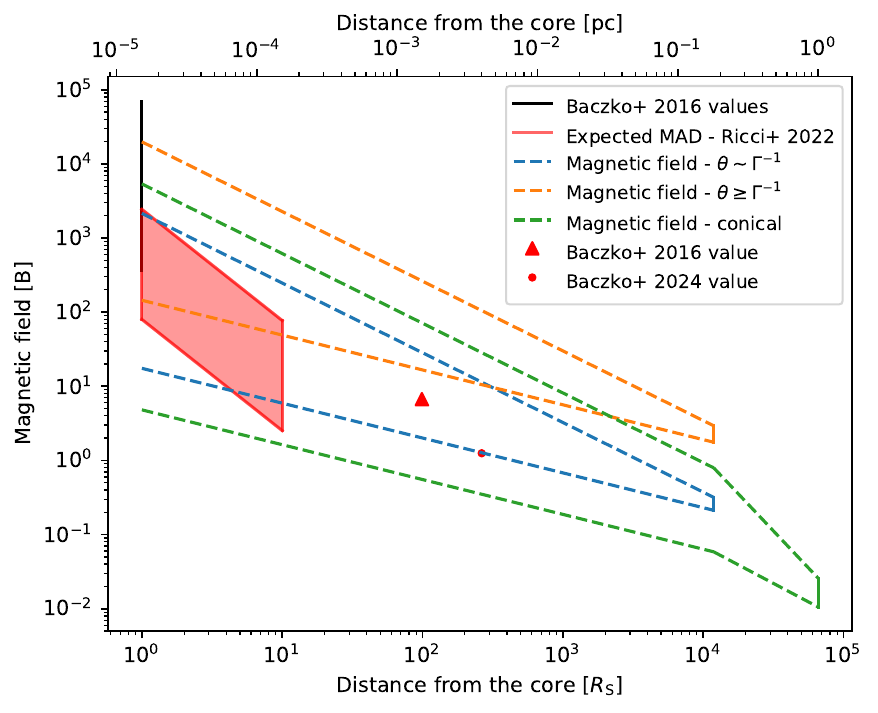}\par
    \caption{Average spatial magnetic field evolution for the two jets in NGC\,1052. The black horizontal line represents the magnetic field strength estimations \citet{Baczko2016} at $1\,\mathrm{R_S}$, while the green, blue, and orange vertical lines are the values (with their uncertainties) from Equations \ref{eq:B_field}, \ref{eq:B_field_2}, and \ref{eq:B_field_4}, respectively. The dashed lines are the extrapolation assuming $b = 0.47$ and $b = 0.94$ ($B \propto \mathrm{z}^{-b}$) for the parabolic region and $b = 1.0$ and $b = 2.0$ for the conical one. The red triangle is a lower limit estimation for the magnetic field from \citet{Baczko2016}, while the red dot is from \citet{Baczko2024}. The red box represents the expected magnetic field needed to saturate the accretion disk.}
    \label{fig:bfield}
\end{figure}

\paragraph{Saturation magnetic field.}

First, we calculate the magnetic field strength along the jet, extrapolate it to its base, and then compare it with the expected magnetic field needed to saturate the disk and form a MAD.
To calculate the magnetic field, we employ the equations reported in \citet{Lobanov1998, Ricci2022, Nokhrina2024}, in which the field strength on (sub-)parsec scales is calculated starting from the knowledge of the core-shift through the parameter:
\begin{equation}
    \Omega_{r \nu} = 4.85 \times 10^{-9} \frac{\Delta r_{\mathrm{mas}} D_\mathrm{L}}{(1 + z)^2} \Bigg(\frac{\nu_1^{1/k_r} \nu_2^{1/k_r}}{\nu_2^{1/k_r} - \nu_1^{1/k_r}}\Bigg) \ \mathrm{pc} \ \mathrm{GHz}^{1/k_r} \, ,
    \label{eq:O_rnu}
\end{equation}
in which $\Delta r_{\mathrm{mas}}$ is the core shift between the two frequencies $\nu_1$ and $\nu_2$, and $D_\mathrm{L}$ is the luminosity distance of the source.

By assuming a jet expanding with a conical shape and $k_r = 1$, the magnetic field at one parsec from the core can be computed as \citep{Lobanov1998}: 
\begin{equation}
    B_1 = 0.025 \ \Bigg[\frac{\Omega_{r \nu}^3 (1 + z)^3}{\delta^2 \phi \mathrm{sin}^2\theta}\Bigg]^{1/4} \ \mathrm{G} \, ,
    \label{eq:B_field}
\end{equation}
where $\delta$ is the Doppler factor, and $\phi$ is the intrinsic jet half-opening angle.

To compute $\Omega_{r \nu}$, we employ the core shift between 1 and 5\,GHz, as these frequencies correspond to a displacement close to the distance of one parsec.
Using the values reported in \citet{Baczko2022}, we obtain $\Omega_{r \nu} = 1.01 \pm 0.10 \, \mathrm{pc} \, \mathrm{GHz}$.
For the jet parameters, we consider average values between the eastern and western jets.
We assume a jet speed of $\beta = 0.40 \, \pm \, 0.20$, consistent with the average values extrapolated in \citet{Baczko2016} and the speed of components closer to one parsec observed in this work, as shown in Sect.\,\ref{sec:modelfit_kinematics}.
The jet half-opening angle at one parsec is $\phi = (4.0 \pm 2.0)\degree$ \citep{Baczko2022}. 
Overall, with such parameters, we obtained an average magnetic field strength of $B_1 = (0.018 \pm 0.008) \, \mathrm{G}$ at one parsec on both jets.

However, as shown in Sects.\,\ref{sec:jetorigin} and \ref{sec:collimation}, $k_r$ is lower than one and the jets of NGC\,1052 show a quasi-parabolic geometry up to a few mas.
Therefore, for a more precise estimation of the magnetic fields we need to drop the assumption of a conical jet used to compute Eq.\,\ref{eq:B_field}.
To do so, we use the model shown in \citet{Ricci2022, Nokhrina2024}, which allows us to compute the magnetic field strength starting from the core-shift parameters in a generic jet geometry.
There are three different cases, depending on the interplay between the viewing angle and the inverse of the Lorentz factor, namely $\theta \sim \Gamma^{-1}$, $\theta \leq \Gamma^{-1}$, and $\theta \geq \Gamma^{-1}$.
For NGC\,1052 the first and third scenarios are the most plausible ones, since $\theta \geq 80\degree \geq 1.1 \, \mathrm{rad}$ and $\Gamma^{-1} \sim 0.92$ (using an average $\beta \sim 0.40$, see Table \ref{tab:jet_speed}).
In the first scenario, $\theta \sim \Gamma^{-1}$:
\begin{equation}
    B_\mathrm{z} = 0.025 \ \Bigg[\frac{\Omega_{r \nu}^{6 \psi} (1 + z)^3}{\mathsf{z}^{6 \psi}r_\mathsf{z}\delta^2 \mathrm{sin}^{6 \psi -1}\theta}\Bigg]^{1/4} \ \mathrm{G} \, , 
    \label{eq:B_field_2}
\end{equation}
while in the third, $\theta \geq \Gamma^{-1}$:
\begin{equation}
    B_\mathrm{z} = 0.025 \ \Bigg[\frac{\Omega_{r \nu}^{8 \psi} (1 + z)^3 (1-\cos\theta)^2}{\mathsf{z}^{8 \psi}r_\mathsf{z} \mathrm{sin}^{8 \psi-1}\theta}\left(\frac{r_\mathsf{z}}{R_L}\right)^2\Bigg]^{1/4} \ \mathrm{G} \, .
    \label{eq:B_field_4}
\end{equation}
In such equations, $r_\mathsf{z}$ is the jet radius (in pc) at a distance $\mathrm{z}$ (in pc) from the radio core, $\psi$ is the power-law index of the jet expansion profile ($d = \mathrm{z}^{\psi}$), and $R_L$ is the light cylinder radius (in pc).
The latter is the intrinsic length scale in MHD models and is expressed as $R_\mathrm{L} = r_{\mathrm{br}} / d_*(\sigma_\mathrm{M})$ 
where $r_{\mathrm{br}}$ is the jet radius at the transition distance and $d_*$ is the non-dimensional jet radius whose value depends on the Michel magnetization $\sigma_\mathrm{M}$ \citep[see][and references therein]{Beskin2017}.

We compute Equations \ref{eq:B_field_2} and \ref{eq:B_field_4} for the transition distance, for which we assume the average between the two jets according to Sect.\,\ref{sec:collimation}, namely $\mathrm{z}_\mathrm{br} = (0.18 \pm 0.02) \, \mathrm{pc}$. 
The corresponding average jet radius is $\mathrm{r}_\mathrm{br} = (0.011 \pm 0.003) \, \mathrm{pc}$.
As a power-law index for the jet expanding profile, we assume the average between the upstream values in the two jets, namely $\psi = 0.47 \pm 0.04$.
The Michel magnetization parametrizes the terminal Lorentz factor reached by the jet flow when all the electromagnetic flux is transferred to the particles \citep[see, for example][and references therein]{Nokhrina2015}.
The highest speed observed in NGC\,1052 is $\beta \sim 0.65$ \citep{Baczko2016, Baczko2019} leading to $\sigma_\mathrm{M} = \Gamma \sim 1.32$.
In \citet{Nokhrina2019}, the different values of $d_*(\sigma_\mathrm{M})$ are given starting from a magnetization of five, which we therefore assume as our lower limit.
However, because of the large jet viewing angle, the radio emission detected from NGC\,1052 is most likely associated with the slower sheath, prohibiting us to detect the faster central spine.
NGC\,1052 is a low-power radio galaxy, and its aligned counterpart would correspond to a BL Lac object, with expected bulk Lorentz factors of $\Gamma \sim 10$ \citep[see, for example,][]{Hovatta2009}, leading to a magnetization $\sigma_\mathrm{M} = 10$. 
As a consequence, we calculate Eq.\,\ref{eq:B_field_4} using values of $d_*(\sigma_\mathrm{M})$ assuming magnetizations of five and ten, and then averaging the result.
We constrain a magnetic field strength at the transition distance of $B = 0.27 \pm 0.05 \, \mathrm{G}$ and $B = 2.35 \pm 0.59 \, \mathrm{G}$, using Equations \ref{eq:B_field_2} and \ref{eq:B_field_4}, respectively.
While the former is in agreement with the results from Eq.\,\ref{eq:B_field}, the latter is roughly one order of magnitude higher. 
The obtained magnetic field strengths are shown in Fig.\,\ref{fig:bfield}.

The next step is to extrapolate the magnetic field strength computed downstream of the jets up to their onset.
To do that, we need to make assumptions on the radial magnetic field dependence, namely $b$ in $B(\mathrm{z}) \propto \mathrm{z}^{-b}$.
When the magnetic fields in jets are dominated by their toroidal component, $b$ is equal to the inverse of the power-law index for the jet width, namely $-\psi$.
On the contrary, when the magnetic fields are dominated by their poloidal component, $b = -2\psi$ \footnote{Despite $b$ being the opposite of $\psi$, because of the convention $B(\mathrm{z}) \propto \mathrm{z}^{-b}$, $b$ will have a positive sign.}.
Having no constraints on the magnetic field geometry, we extrapolate the magnetic field strength considering both cases using $\psi = 0.47$.
To include all possible magnetic field values and account for uncertainties, we extrapolate the lower limits of the magnetic field estimations (value minus the error) with $b = 0.47$ and the upper limits with $b = 0.94$ ($b = 1$ and $b = 2$ for the conical region, respectively).

In the following, we compare our estimated $B$ with previous works. 
Namely, in \citet{Baczko2016} the authors inferred a lower limit on the magnetic field strength of $6.7\, \mathrm{G}$ at $1.5 \, \mathrm{mpc} = 99\,\mathrm{R_S}$ by means of synchrotron cooling arguments which leads to extrapolate values of $360 \, \mathrm{G} \leq B \leq 6.9 \times 10^4 \, \mathrm{G}$ at $1\,\mathrm{R_\mathrm{S}}$.
In their work, they assumed the decrease in radio emission with the distance from the core at 86\,GHz only to the cooling, without effects from the free--free absorption.
This hypothesis is supported by what is proposed in this work.
Additionally, in \citet{Baczko2024} the authors constrained a magnetic field strength of $B \sim 1.25 \, \mathrm{G}$ at $43\,\mathrm{\mu as} = 264\,\mathrm{R_S}$ by fitting the synchrotron-self absorbed spectrum of the core component between 22\,GHz to 230\,GHz.
The literature data points are consistent with the magnetic field strengths calculated using Eqs.\,\ref{eq:B_field} and \ref{eq:B_field_2} when evolving with $b \geq 0.47$, with the values at $1\,\mathrm{R_\mathrm{S}}$ in agreement with our estimations with a magnetic field geometry of $b \sim 1.0$.

Merging all the information together, as shown in Fig.\,\ref{fig:bfield} a magnetic field evolving with an index $0.47 \leq b \leq 0.94$ is expected, since none of the magnetic field estimations along the jet match the values at $1\,\mathrm{R_\mathrm{S}}$ when evolving with $b \sim 0.47$.
However, the situation is more complicated.
As shown in Fig.\,\ref{fig:coll} on scales smaller than $10^4\,\mathrm{R_S}$, the parabolic profiles inferred are the average jet geometries, with the jet alternating regions of steeper and flatter expansion rates.
Since the magnetic field evolution follows the same trend ($b \propto -\psi$), changes in the jet width profile reflect into changes in the magnetic field one.
In particular, we observe regions where the jet expansion profile nearly saturates ($\psi \sim 0$), such as near the jet injection point, where, consequently, the magnetic field strength remains approximately constant as well.
As a consequence, to reach magnetic field strengths in agreement with the findings of \citet{Baczko2016} at $1\,\mathrm{R_\mathrm{S}}$, when the jet is expanding the magnetic field need to evolve with an index close to one. 
A magnetic field trend with the distance $b \sim 1$ has been recovered for M\,87 and NGC\,315 as well on similar scales \citep[see][respectively]{Ro2023, Ricci2025} and is in agreement with previous findings for the source \citep{Baczko2016}.
Values of $b \sim 1$ can be a consequence of a purely toroidal magnetic field being highly dissipated to accelerate the jet or a magnetic field with an important contribution from its poloidal component leading to a helical field geometry up to scales of $\sim 10^4\,\mathrm{R_S}$ at least.

Finally, we can compare the magnetic field strength in the nuclear region with the magnetic field needed to saturate the disk (precisely, magnetic force $\geq$ gravitational force) and form a MAD.
To compute the saturation fields, we use Eq.\,17 in \citet{Ricci2022}.
For the mass accretion rate ($\dot{M} = L_\mathrm{acc} / \eta c^2$), we assume the value of $L_\mathrm{acc} = 10^{41.24} \, \mathrm{erg \, s^{-1}}$ from X-ray measurements \citep{Woo2002,Gonzalez2009}.
As radiative efficiencies, we assume $\eta = 0.001, 0.01,$ and $0.4$.
On the one hand, NGC\,1052 has been proposed to have an ADAF, which is a type of disk with efficiencies lower than the standard thin accretion disks \citep{Yuan2014} and that can go down to $\eta = 0.001, 0.01$ \citep{Narayan1995, Mahadevan1997}. 
On the other hand, $\eta = 0.4$ has been assumed for NGC\,1052 in sample studies including it, and we consider it for consistency \citep[][]{Zamaninasab2014}.
Finally, the radial velocity of the gas is assumed to be closer to the speed of light and the scale height of the disk $h/r = 0.25 \pm 0.10$.
This value and uncertainty is chosen to include the different possible black hole spin scenarios, as shown in Fig.\,7 in \citep{Narayan2022}.
Considering all the possible sources of uncertainties and the different values of $\eta$, we find the saturation magnetic field to vary between $80\,\mathrm{G} \leq B \leq 2438\,\mathrm{G}$ at $1\,\mathrm{R_\mathrm{S}}$ and $3\,\mathrm{G} \leq B \leq 77\,\mathrm{G}$ at $10\,\mathrm{R_\mathrm{S}}$, chosen as an arbitrary upper limit on the putative magnetosphere.
Such ranges are shown by the red box in Fig.\,\ref{fig:bfield}.
Following this calculation, the magnetic field strength required to saturate the disk and reach the MAD state is achieved when $b \gtrsim 0.47$, as already suggested for NGC\,1052 earlier in this section.
This analysis provides a very first indication that the accretion disk in NGC\,1052 might have reached a MAD state.
Overall, our extrapolated magnetic field strength with $b \sim 1$ would lead to values at the jet base in the range of $\sim 10^2-10^4 \, \mathrm{G}$, consistent with the findings for other radio galaxies, such as NGC\,315 \citep{Ricci2022, Ricci2025} and M\,87 \citep{Ro2023}.
However, in the scenario the real magnetic field strengths at the jet base are in the range of tens of Gauss, such as the ones proposed for M\,87 from EHT observations \citep{eht_bfield}, values of $b \sim 0.50$ would best describe the field strength extrapolation.
In this case, either the magnetic field evolves up to $1\, R_\mathrm{S}$ with a purely toroidal configuration or the fields saturate more downstream the jet, forming a magnetosphere on scales of tens of $R_\mathrm{S}$, as shown in Fig.\,\ref{fig:bfield}.

\paragraph{Magnetic flux}

An additional approach to assess whether the accretion disk in NGC\,1052 has reached a MAD state involves comparing the magnetic flux along the jet with the one expected from a MAD.
This comparison can be done following the work by \citet{Zamaninasab2014}.
In particular, under the assumption that the gravitational potential of the black hole is the main regularizer of the width of the optical broad lines coming from the clouds surrounding the central region, the expected MAD flux is:
\begin{equation}
    F_\mathrm{MAD} = 2.4 \times 10^{34} \Bigg[ \frac{\eta}{0.4} \Bigg]^{-1/2} \Bigg[ \frac{M_\mathrm{BH}}{10^9 M_\odot} \Bigg] \Bigg[ \frac{L_\mathrm{acc}}{1.26 \times 10^{47} \, \mathrm{erg/s}} \Bigg]^{1/2} \, \mathrm{G} \, \mathrm{cm^2}.
\label{eq:expectedMAD}
\end{equation}
We use the same parameters as for calculating the saturation fields, namely $L_\mathrm{acc} = 10^{41.24} \, \mathrm{erg \, s^{-1}}$ and $\eta = 0.001, 0.01,$ and $0.4$ obtaining, respectively $F_\mathrm{MAD} = 7.45 \times 10^{31} \, \mathrm{G} \, \mathrm{cm^2}$, $F_\mathrm{MAD} = 2.35 \times 10^{31} \, \mathrm{G} \, \mathrm{cm^2}$, and $F_\mathrm{MAD} = 3.72 \times 10^{30} \, \mathrm{G} \, \mathrm{cm^2}$.

To calculate the poloidal magnetic flux in the jet, we use Eq.\,1 of \citet{Zamaninasab2014}:

\begin{equation}
    F_\mathrm{jet} = 1.2 \times 10^{34}  f(a_*) \Gamma \phi \Bigg[ \frac{M_\mathrm{BH}}{10^9 M_\odot} \Bigg] \Bigg[ \frac{B_1}{1\,\mathrm{G}} \Bigg] \, \mathrm{G} \, \mathrm{cm^2} \, ,
\label{eq:calculatedflux}
\end{equation}
in which $f(a_*) = 1/a_* \big[ 1 + (1-a^2)\big]$, and $a_*$ is the dimensionless spin of the central black hole.
This equation is based on the assumption of a jet expanding conically and allows for calculating the magnetic field flux at one parsec downstream of the jet.
While we know that the assumption of a conical jet does not hold for NGC\,1052 (it does at one parsec, but not upstream, see Sect.\,\ref{sec:collimation}), we use this equation to extrapolate an order of magnitude for the magnetic flux.
Being the spin of the black hole unknown, we use the two opposite scenarios of a slowly and highly rotating object, namely $a_* = 0.10$ and $a_* = 0.99$, obtaining $F_\mathrm{jet} = (3.01 \pm 1.32) \times 10^{32} \, \mathrm{G \, cm^2}$ and $F_\mathrm{jet} = (1.74 \pm 0.8) \times 10^{31} \, \mathrm{G \, cm^2}$, respectively.

Assuming magnetic flux conservation from one parsec up to the jet injection point, the calculated flux via Eq.\,\ref{eq:calculatedflux} is consistent with the value expected in the MAD scenario under the conditions of low radiative efficiency (in agreement with the disk being an ADAF) and a rapidly spinning central supermassive black hole.
Interestingly, the nuclear region of 3C\,84 is also compatible with this model, perhaps indicative of such conditions being prevalent in nearby radio galaxies \citep{Paraschos2023, Paraschos2024a}.


\section{Summary} \label{sec:Conclusions}

We have analyzed new 43\,GHz and 86\,GHz observations of NGC\,1052 to investigate the symmetry of its jets and their magnetic field properties.
Our results can be summarized as follows.

\begin{itemize}
    \item We obtained three new images at 43\,GHz, which show a morphology consistent with the most recent epochs presented in \citet{Baczko2019}, characterized by the absence of a bright core component in correspondence with the jet launching region. 
    Additionally, the images reveal asymmetries between the eastern and western jets, as well as a time-variable morphology, likely associated with the passage of components downstream in both jets.
    \item Due to various issues during the 2021 observations, we were only able to acquire a single new image at 86\,GHz. 
    Using a stacked 86\,GHz image, combining our new data with the map published in \citet{Baczko2024}, we derived the flux density profile of both jets, finding a high degree of symmetry. 
    \item Thanks to the new observations presented here, we updated the jet width profile published in \citet{Baczko2022, Baczko2024}. 
    We infer symmetric, parabolic jets on sub-mas scales with the eastern one transitioning to a quasi-conical shape at $\mathrm{z_b}~=~1.2 \pm 0.3 \, \mathrm{mas}$ while the western one at $\mathrm{z_b}~=~2.7~\pm~0.2 \, \mathrm{mas}$. 
    Downstream, the jets present asymmetries, with the receding one showing a steeper jet geometry.
    
    \item 
    The (43-86)\,GHz spectral index map indicates that the core region is highly inverted, with a spectral index as high as $\alpha_\mathrm{thick} \sim 3.3$, suggesting that free--free absorption still impacts the emission at 43\,GHz. 
    This scenario is further supported by the flux density profile at 43\,GHz, where the western jet shows a brightness approximately three times lower than the eastern jet counterpart at the same distance.
    \item Overall, we propose that the bulk flows are intrinsically symmetric when launched.
    The asymmetries observed at 43\,GHz are likely a combination of the free--free absorption exercised by the clumpy torus on the western jet and the propagation of jet components downstream of the two jets.
    This scenario is consistent with the numerical simulations presented in \citet{Saiz2025} for NGC\,1052, namely, from symmetrically launched jets, time-dependent asymmetries might arise due to the passage of the jet perturbations.
    \item Finally, we explore the magnetic field strength along the jets in NGC\,1052 by employing our knowledge of the core shift and the extent of the collimation region in both jets. 
    Our analysis is consistent with previous estimates of the magnetic field. 
    This field strength is expected to be strong enough to saturate the central region, implying the possible presence of a MAD.
    This scenario is further examined by analyzing the magnetic flux at the jet base. 
    The magnetic flux expected in a MAD scenario aligns with the jet flux in the scenario of low radiative efficiency and highly spinning central supermassive black hole.
\end{itemize}

In conclusion, the new datasets presented in this study provide important insights into the jet properties of NGC 1052, from the role of the absorbing torus to the intrinsic jet symmetry and their magnetic field content.
Future multi-frequency and simultaneous observations will be essential to further constrain these properties and explore the proposed scenarios in greater detail.

\begin{acknowledgements} 
We would like to thank the referee for the meaningful comments, which improved the manuscript. This research is funded by the Deutsche Forschungsgemeinschaft (DFG, German Research Foundation) – project number 443220636 and by the European Research Council advanced grant “M2FINDERS - Mapping Magnetic Fields with INterferometry Down to Event hoRizon Scales” (Grant No. 101018682).
SdP gratefully acknowledges funding from the European Research Council (ERC) under the European Union's Horizon 2020 research and innovation programme (grant agreement No 789410, PI: S. Aalto).
MP acknowledges support from \texttt{MICIU/AEI/10.13039/501100011033} and FEDER, UE, via the grant PID2022-136828NB-C43, from the Generalitat Valenciana through grant CIPROM/2022/49, and from the Astrophysics and High Energy Physics project program supported by the Spanish Ministry of Science and Generalitat Valenciana with funding from European Union NextGenerationEU (\texttt{PRTR-C17.I1}) through grant \texttt{ASFAE/2022/005}. 
This research has made use of data obtained with the Global Millimeter VLBI Array (GMVA), which consists of telescopes operated by the MPIfR, IRAM, Onsala, Metsahovi, Yebes, the Korean VLBI Network, the Greenland Telescope, the Green Bank Observatory and the Very Long Baseline Array (VLBA). The VLBA and the GBT are facilities of the National Science Foundation operated under cooperative agreement by Associated Universities, Inc. The data were correlated at the correlator of the MPIfR in Bonn, Germany.
Partly based on observations with the 100-m
telescope of the MPIfR (Max-Planck-Institut für Radioastronomie) at Effelsberg.
This work made use of Astropy:\footnote{http://www.astropy.org} a community-developed core Python package and an ecosystem of tools and resources for astronomy \citep{astropy:2013, astropy:2018, astropy:2022}.
\end{acknowledgements}

\bibliographystyle{aa.bst}
\bibliography{bibliography}

\begin{appendix}

\section{Problems afflicting the observations} \label{app:problems}

Concerning the 3\,mm observations, in April 2021, Metsahovi and Pico Veleta were partially afflicted by poor weather, Yebes participated in only a few scans and fringes were detected only for one of them, Green Bank had pointing problems, and the VLBA antennas suffered from problems at the 3\,mm receiver \footnote{ \url{https://www3.mpifr-bonn.mpg.de/div/vlbi/globalmm/sessions/apr21/feedback_apr21.asc}}.
The October 2021 session suffered from poor weather conditions at several stations (mostly Europeans, such as Effelsberg, Onsala, and Metsahovi), no fringes at Yebes, the data from Green Bank were corrupted and so discarded, and an unusually large number of scans with high system temperature ($> 1000 \, \mathrm{K}$) at the VLBA stations\footnote{\url{https://www3.mpifr-bonn.mpg.de/div/vlbi/globalmm/sessions/oct21/feedback_oct21.asc}}.
The aforementioned problems, combined with the relatively low flux density of NGC~1052 and consequent high uncertainties in the final calibrated data, did not allow us to obtain reliable 86\,GHz images from those two observing sessions.

At 43\,GHz, the observations performed in April and October 2021 suffered from flux scaling problems \citep[the same problems were reported in][]{Ricci2025}.
In April 2021, the output from the correlator at 7\,mm contains all autocorrelation records twice, leading to a flux density lower than expected.
To account for this issue, the baseline-based fluxes were multiplied by a factor of $\sqrt{2}$, following the suggestions provided by the correlator team. 
For the October 2021 observations, the gain curves originally provided from the VLBA team were corrupted due to issues with the focus and rotation at the VLBA stations
\footnote{\url{https://science.nrao.edu/facilities/vlba/data-processing/7mm-performance-2021}}. 
This problem was fixed by applying the updated gain curves provided by the VLBA staff to the data.

\section{Modelfit components and best-fit tables} \label{app:model}

\begin{figure*}[t]
    \centering
    \begin{multicols}{2}
    \includegraphics[width=0.9\linewidth]{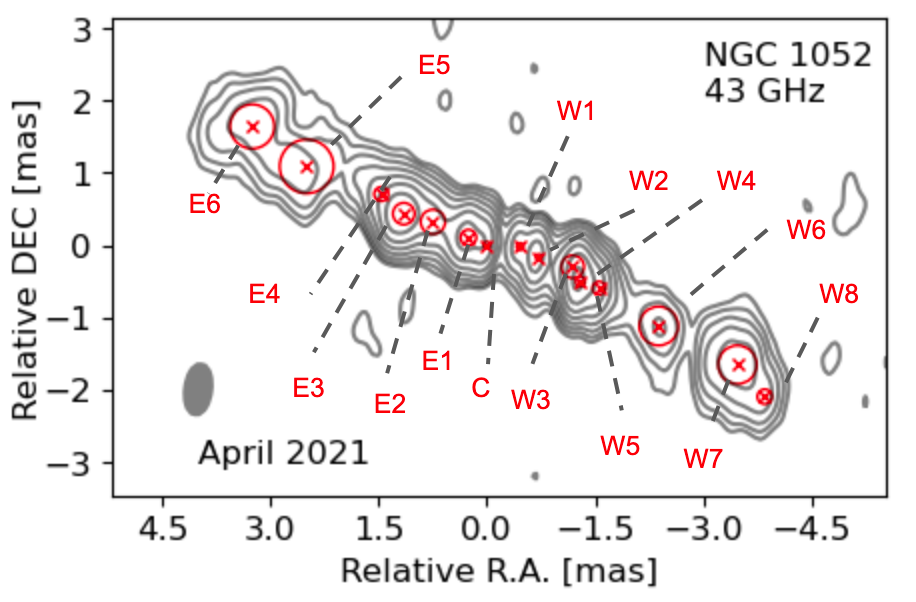}\par
    \includegraphics[width=0.9\linewidth]{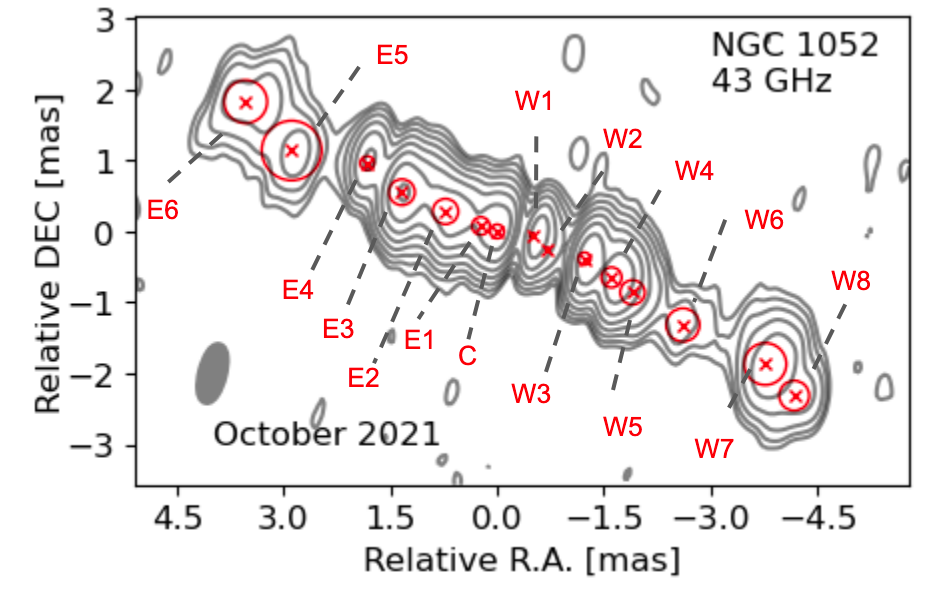}\par
    \end{multicols}
    \begin{multicols}{2}
    \includegraphics[width=0.9\linewidth]{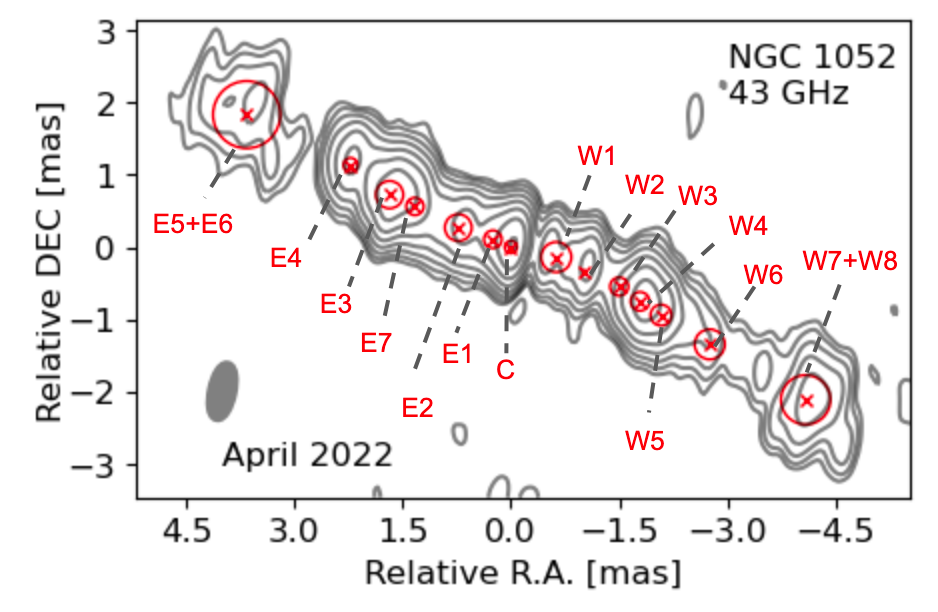}\par
    \includegraphics[width=\linewidth]{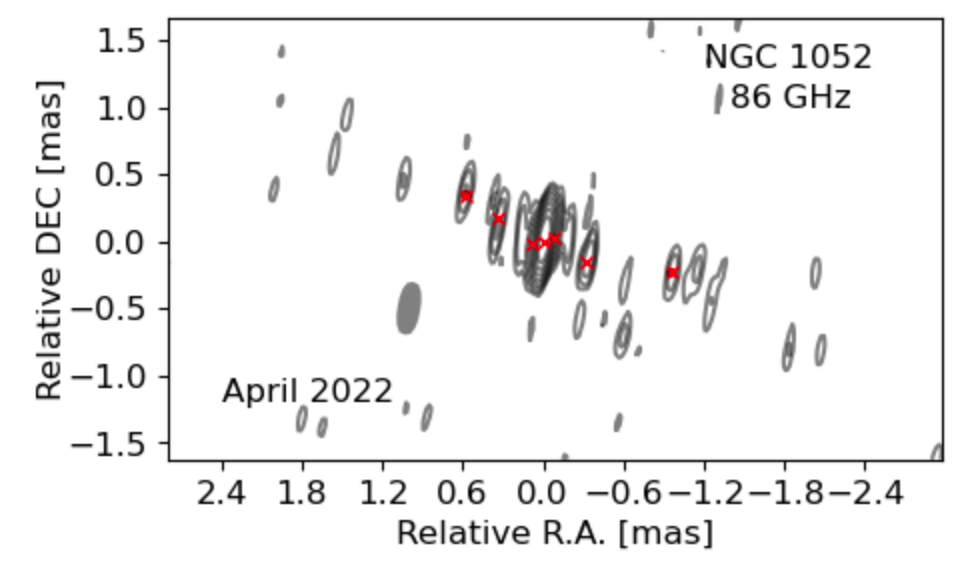}\par
    \end{multicols}
    \caption{Modelfit components (in red) overlaid to the images presented in this paper. Upper panels: 43\,GHz April 2021 (left), 43\,GHz October 2021 (right). Lower panels: 43\,GHz April 2022 (left), 86\,GHz April 2022 (right).}
    \label{fig:modelfit_comps}
\end{figure*}

In this section, we report the main parameters of the modelfit components in Table \ref{tab:modelfit} together with the images with the modelfit component overlaid in Fig.\,\ref{fig:modelfit_comps}.
Additionally, we report the best-fit values (Table \ref{tab:bestfit}) for the different fits shown in Fig.\,\ref{fig:coll}.

\begin{table}[h!]
\begin{minipage}{0.5\textwidth}
\centering
\caption{Modelfit parameters of the Gaussian components at 43\,GHz and after the double-line, at 86\,GHz.}
\label{tab:modelfit}
\setlength{\tabcolsep}{3pt} 
\begin{tabular}{ccccccc}
\hline
\begin{tabular}[c]{@{}c@{}}Obs.\\ date\end{tabular} &
\begin{tabular}[c]{@{}c@{}}S\\ {[}mJy{]}\end{tabular} &
\begin{tabular}[c]{@{}c@{}}d\\ {[}mas{]}\end{tabular} &
\begin{tabular}[c]{@{}c@{}}z\\ {[}mas{]}\end{tabular} &
\begin{tabular}[c]{@{}c@{}}PA\\ {[}deg{]}\end{tabular} &
\begin{tabular}[c]{@{}c@{}}$\theta_\mathrm{lim}$\\ {[}mas{]}\end{tabular} &
\begin{tabular}[c]{@{}c@{}}Comp.\\ name\end{tabular}
\\ \hline

A. 21
 & 12.18 & 0.205 & $-$4.39 & 61.38 & 0.03 & W8 \\
 & 59.96 & 0.533 & $-$3.85 & 64.48 & 0.013 & W7 \\
 & 19.85 & 0.539 & $-$2.63 & 64.78 & 0.023 & W6 \\
 & 45.32 & 0.183 & $-$1.66 & 69.0 & 0.015 & W5 \\
 & 104.03 & 0.132 & $-$1.39 & 68.65 & 0.01 & W4 \\
 & 42.96 & 0.315 & $-$1.23 & 75.82 & 0.016 & W3 \\
 & 34.64 & 0.081 & $-$0.74 & 76.89 & 0.018 & W2 \\
 & 16.92 & 0.096 & $-$0.46 & 87.45 & 0.025 & W1 \\
 & 42.46 & 0.093 & 0.0 & 0.0 & 0.016 & C \\
 & 86.12 & 0.218 & 0.28 & $-$112.11 & 0.011 & E1 \\
 & 58.17 & 0.339 & 0.81 & $-$113.31 & 0.014 & E2 \\
 & 149.56 & 0.301 & 1.23 & $-$110.67 & 0.008 & E3 \\
 & 21.41 & 0.196 & 1.63 & $-$115.77 & 0.022 & E4 \\
 & 44.16 & 0.747 & 2.72 & $-$113.58 & 0.016 & E5 \\
 & 38.42 & 0.613 & 3.64 & $-$116.8 & 0.017 & E6 \\

\hline

O. 21 & 13.36 & 0.421 & $-$4.77 & 61.03 & 0.031 & W8 \\
 & 72.4 & 0.591 & $-$4.2 & 63.66 & 0.013 & W7 \\
 & 11.11 & 0.468 & $-$2.92 & 63.24 & 0.034 & W6 \\
 & 38.12 & 0.346 & $-$2.09 & 65.74 & 0.018 & W5 \\
 & 127.35 & 0.284 & $-$1.73 & 68.09 & 0.01 & W4 \\
 & 113.69 & 0.176 & $-$1.28 & 72.65 & 0.011 & W3 \\
 & 15.28 & 0.056 & $-$0.75 & 70.43 & 0.029 & W2 \\
 & 39.7 & 0.069 & $-$0.5 & 83.9 & 0.018 & W1 \\
 & 64.2 & 0.197 & 0.0 & 0.0 & 0.014 & C \\
 & 95.37 & 0.251 & 0.24 & $-$108.6 & 0.012 & E1 \\
 & 161.21 & 0.359 & 0.78 & $-$111.2 & 0.009 & E2 \\
 & 206.4 & 0.365 & 1.45 & $-$112.44 & 0.008 & E3 \\
 & 36.14 & 0.198 & 2.06 & $-$117.75 & 0.019 & E4 \\
 & 34.75 & 0.84 & 3.1 & $-$111.53 & 0.019 & E5 \\
 & 33.16 & 0.604 & 3.98 & $-$117.33 & 0.02 & E6 \\

\hline

A. 22 & 57.1 & 0.685 & $-$4.6 & 62.61 & 0.014 & W7+W8 \\
 & 17.09 & 0.418 & $-$3.07 & 63.93 & 0.025 & W6 \\
 & 41.88 & 0.291 & $-$2.28 & 65.6 & 0.016 & W5 \\
 & 90.94 & 0.261 & $-$1.94 & 67.18 & 0.011 & W4 \\
 & 22.34 & 0.254 & $-$1.6 & 69.96 & 0.022 & W3 \\
 & 25.08 & 0.033 & $-$1.07 & 71.64 & 0.021 & W2 \\
 & 35.5 & 0.418 & $-$0.64 & 77.55 & 0.017 & W1 \\
 & 91.0 & 0.171 & 0.0 & 0.0 & 0.011 & C \\
 & 45.75 & 0.247 & 0.27 & $-$113.31 & 0.015 & E1 \\
 & 59.01 & 0.372 & 0.78 & $-$110.29 & 0.013 & E2 \\
 & 45.05 & 0.243 & 1.44 & $-$112.94 & 0.015 & E7 \\
 & 149.53 & 0.381 & 1.83 & $-$113.35 & 0.008 & E3 \\
 & 43.3 & 0.197 & 2.5 & $-$117.09 & 0.016 & E4 \\
 & 37.57 & 0.931 & 4.09 & $-$116.61 & 0.017 & E5+E6 \\

\hline
\hline

A. 21 & 1120 & 0.038 & 0.0 & 0.0 & 0.036 & - \\

\hline

O. 21 & 1811 & 0.035 & 0.0 & 0.0 & 0.032 & - \\

\hline

A. 22 & 9.74 & 0.053 & $-$0.98 & 76.27 & 0.017 & - \\
 & 13.93 & 0.011 & $-$0.36 & 64.3 & 0.014 & - \\
 & 26.81 & 0.014 & $-$0.09 & 108.89 & 0.01 & - \\
 & 620.12 & 0.003 & 0.0 & 0.0 & 0.002 & - \\
 & 28.75 & 0.013 & 0.09 & $-$73.62 & 0.01 & - \\
 & 7.05 & 0.011 & 0.37 & $-$116.91 & 0.02 & - \\
 & 9.33 & 0.058 & 0.67 & $-$120.89 & 0.017 & - \\

\hline

\hline
\end{tabular}
\vspace{0.5em}
\begin{flushleft}
\textbf{Notes.} Column 1: date of the observation; Column 2: flux density in mJy; Column 3: FWHM in mas; Column 4: radial distance in mas after the core-shift correction; Column 5: position angle after the core-shift correction; Column 6: resolution limit; Column 7: associated component name at 43\,GHz. A.\ stands for April and O.\ stands for October.
\end{flushleft}
\end{minipage}
\end{table}

\begin{table*}
\centering
\caption{Fit parameters for the broken power law fits to the jet width for the western and eastern jets shown in Fig.\,\ref{fig:coll}.}
\begin{tabular}{lccccccccc}
\hline
Fit to &$W_0$ & $\sigma_{W_0}$ & $k_\mathrm{u}$ & $\sigma_{k_\mathrm{u}}$ & $k_\mathrm{d}$ & $\sigma_{k_\mathrm{u}}$ & $\mathrm{z}_\mathrm{B}$ & $\sigma_{\mathrm{z}_\mathrm{B}}$ & $\chi^2$ \\
\hline
W jet - all & 0.27 & 0.02 & 0.42 & 0.02 & 1.20 & 0.05 & 2.7 & 0.2 & 2.59 \\
W jet - Baczko et al. 22 &0.26 & 0.02 & 0.21 & 0.05 & 1.22 & 0.05 & 2.66 & 0.22 & 1.91 \\
W jet - high freq & 0.26 & 0.03 & 0.42 & 0.02 & 1.26 & 0.19 & 2.7 & 0.4 & 3.99 \\
E jet - all & 0.22 & 0.03 & 0.50 & 0.02 & 0.75 & 0.01 & 1.10 & 0.23 & 2.35 \\
E jet -  Baczko et al. 22 & 0.24 & 0.02 & 0.22 & 0.06 & 0.80 & 0.01 & 1.40 & 0.18 & 1.40 \\
E jet - high freq & 0.18 & 0.03 & 0.46 & 0.04 & 0.90 & 0.06 & 0.85 & 0.21 & 3.56 \\
\hline
\end{tabular}
\begin{flushleft}
\textbf{Notes.} Column 1: which data the fits describe: W stands for western jet and E for eastern; Column 2: initial jet width; Column 3: error on the initial jet width; Column 4: upstream index; Column 5: error on the upstream index; Column 6: downstream index; Column 7: error on the downstream index; Column 8: transition distance; Column 9: error on the transition distance; Column 10: chi square for the fit. 
\end{flushleft}
\label{tab:bestfit}
\end{table*}

\section{86\,GHz alternative image} \label{app:other86}

\begin{figure*}[t]
    \centering
    \begin{multicols}{2}
    \includegraphics[width=\linewidth]{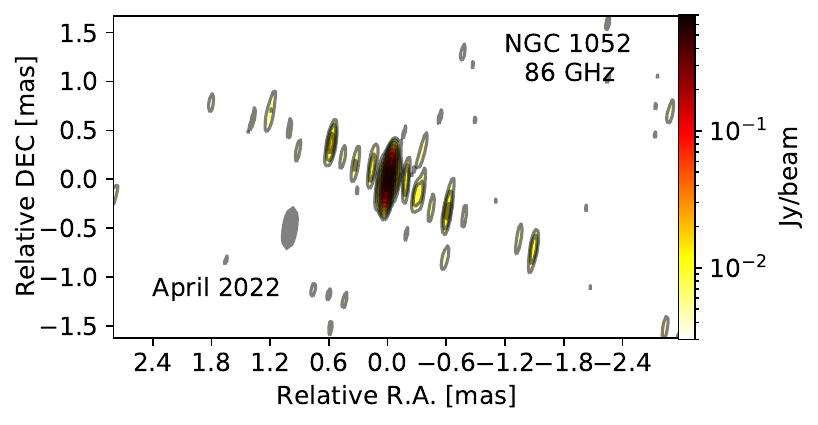}\par
    \includegraphics[width=\linewidth]{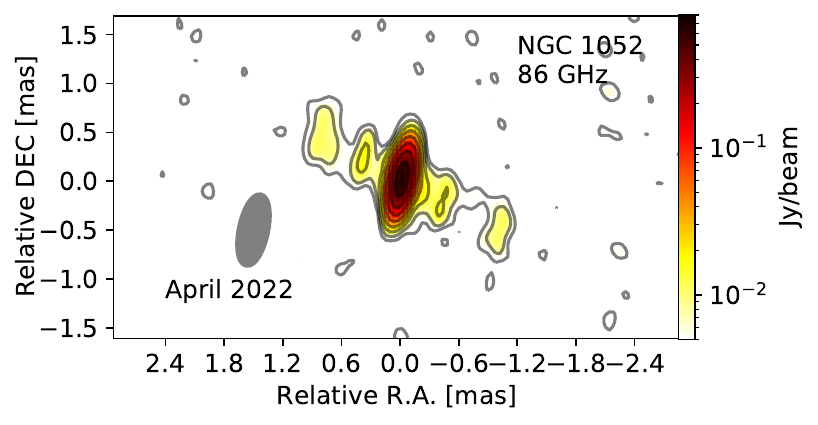}\par
    \end{multicols}

    \caption{86\,GHz image obtained by means of a second and independent imaging run. Left panel: uniform weighting. Right panel: convolved with the 43\,GHz April 2022 beam. The contours start at 3$\sigma$.}
    \label{fig:other86}
\end{figure*}

In Fig.\,\ref{fig:other86}, we report a 3\,mm April 2022 image obtained through an independent imaging run.
The main properties of the resultant image remain largely similar of the one shown in Sect.\,\ref{sec:86GHz}, namely a total flux of $779 \, \mathrm{Jy}$, a brightness peak of $0.66 \, \mathrm{Jy/beam}$, and a rms of $1.023 \, \mathrm{mJy/beam}$.
In this case, the image shows an extended structure within $\sim 0.6 \, \mathrm{mas}$ on both sides which is highly symmetric, with one bright component aligned along the same position angle.
Additionally, we note how the component around $-1 \, \mathrm{mas}$ in the western jet is not recovered.

\begin{figure}[t]
    \centering    \includegraphics[width=0.95\linewidth]{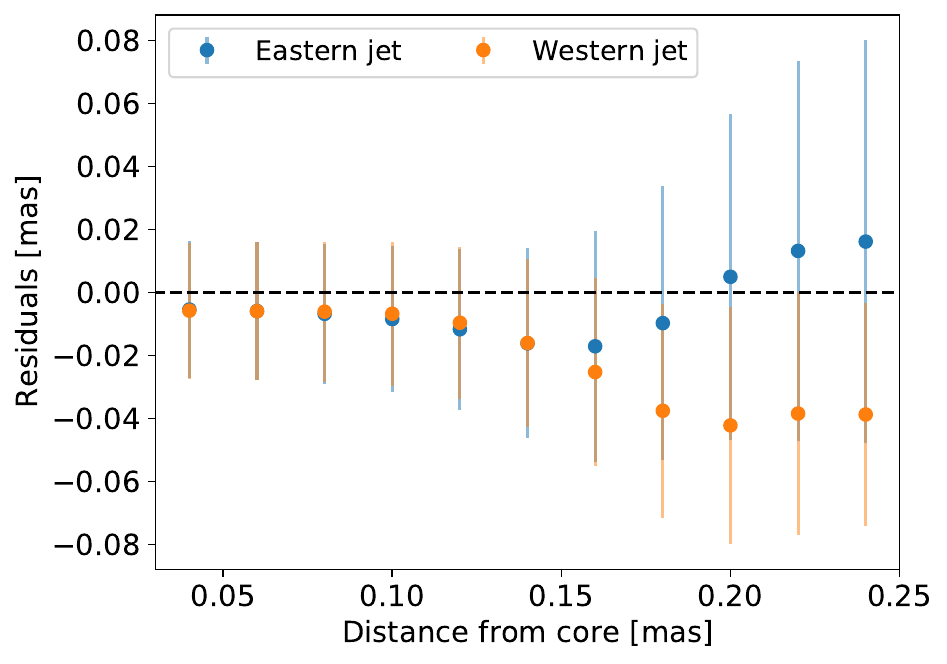}\par
    \caption{Differences in jet width from the two stacked images at 86\,GHz created with the images shown in Fig.\,\ref{fig:86GHzmap} and Fig.\,\ref{fig:other86}, respectively. The differences are consistent with zero within 1$\sigma$ except for the four outermost points in the western jet, which are consistent within 2$\sigma$.}
    \label{fig:residuals}
\end{figure}

We produced an 86\,GHz stacked image using the alternative result shown in this section and derived the jet width profile following the procedure described in Sect.\,\ref{sec:collimation}.
In Fig.\,\ref{fig:residuals} we show the difference between the obtained profile and the one employed in Fig.\,\ref{fig:coll}.
The difference between the vast majority of the data points is consistent with zero within 1$\sigma$, with the exception of the last four points in the western jet, which appear to be slightly thinner and consistent with zero within 2$\sigma$.
Overall, this additional image proves that, despite the slightly different morphology obtained from this cleaning run, the scientific discussion and conclusions we extrapolate from the 86\,GHz data remain unaltered even for a non-univocal solution for imaging.

\section{Stacked images} \label{app:stacked}

\begin{figure*}[t]
    \centering
    \begin{multicols}{2}
    \includegraphics[width=\linewidth]{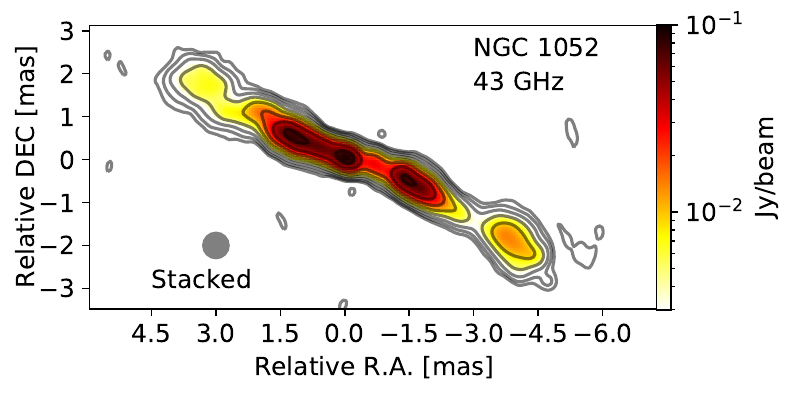}\par
    \includegraphics[width=1.1\linewidth]{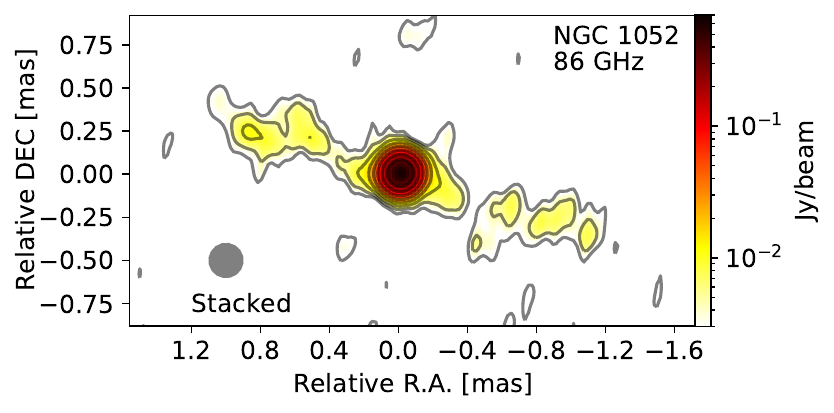}\par
    \end{multicols}
    \caption{Stacked images of NGC\,1052. Left panel: stacked image at 43\,GHz obtained combining the three observations reported in Fig.\,\ref{fig:43GHz_maps}. Each map is aligned with each respective core as described in Sect.\,\ref{sec:jetorigin} and convolved with a circular beam of $0.4 \, \mathrm{mas}$. The contours start at 3$\sigma$ ($\sigma = 0.15 \, \mathrm{mJy/beam}$) and increase by a factor of two up to 384$\sigma$.
    Right panel: stacked image at 86\,GHz obtained by combining the observation reported in Table \ref{table:original_maps} and the 86\,GHz map shown in \citet{Baczko2022}. The circular beam is $0.144 \, \mathrm{mas}$ and the contours start at 3$\sigma$ ($\sigma = 0.8 \, \mathrm{mJy/beam}$).}
    \label{fig:stacked}
\end{figure*}

In this section, we present the two stacked images in Fig.\,\ref{fig:stacked}.
The stacked image at 43\,GHz (left panel) is created by centering the single-epoch observations on their respective core components and by convolving them with the minimum common circular beam of 0.4 mas.
The one at 86\,GHz (right panel) is obtained by combining the April 2022 observation presented here with the 2017 one published in \citet{Baczko2024} with a convolved beam of 0.144 mas.

\section{Alternative spectral index maps} \label{app:othersm}

In this section, we present the collection of the alternative spectral index maps.

\begin{figure}[t]
    \centering
    \includegraphics[width=\linewidth]{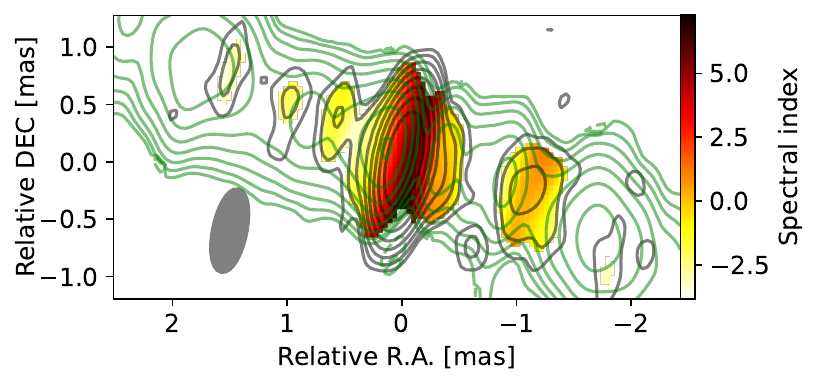}\par
    \caption{Spectral index maps of NGC\,1052 between 43 and 86\,GHz obtained using the shift from \citet{Baczko2024}.}
    \label{fig:sm_anne}
\end{figure}

At first, we present the (43-86)\,GHz spectral index map obtained employing the core shift reported in \citet{Baczko2024} of 0.098 mas in declination and 0.176 in right ascension, namely 2 and 4 pixels, respectively.
The resultant map is shown in Fig.\,\ref{fig:sm_anne}, in which the color scale highlights the spectral index values, the black contours trace the 86\,GHz image, and green contours the 43\,GHz one.
The two images at the two frequencies appear not to be correctly aligned.
On the one hand, the spectral index reaches at the onset of the western jet values up to $\alpha \sim 6.5$, higher than the summed contribution from the self-absorbed synchrotron emission (up to $\alpha \sim 2.5$) and free--free absorption (up to $\alpha \sim 2.0$).
On the other hand, at both frequencies we recover a high-intensity region laying at $\sim -1\ \,\mathrm{mas}$ which appears to be misaligned with the 86\,GHz counterpart, which is too far downstream in the 43\,GHz jet.
Since the core shift is a consequence of the physical properties of the jet, such as the strength of the underlying magnetic fields, variability across different epochs is expected \citep[see, for example,][]{Plavin2019}.

\begin{figure*}
    \centering
    \includegraphics[width=0.49\linewidth]{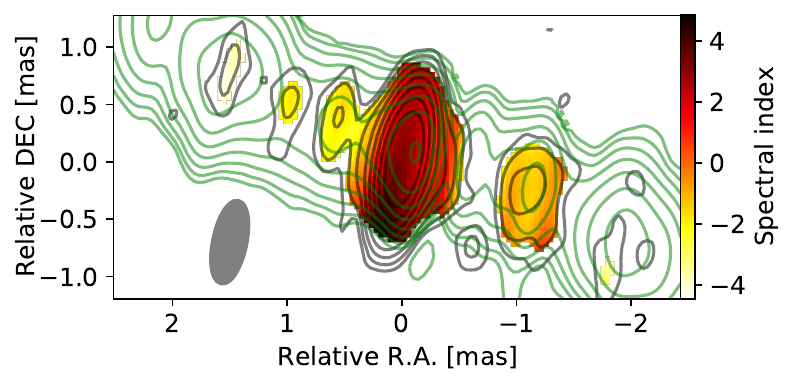}\hfill
    \includegraphics[width=0.49\linewidth]{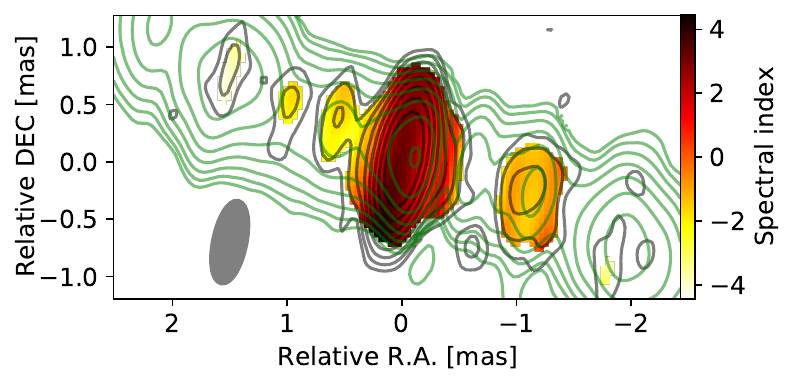}\\
    
    \includegraphics[width=0.49\linewidth]{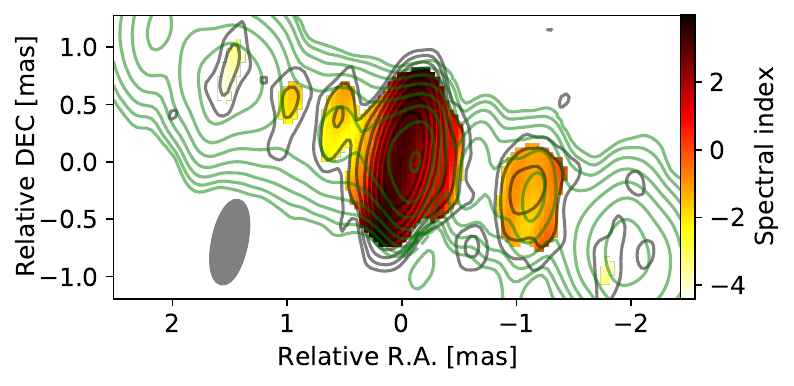}\hfill
    \includegraphics[width=0.49\linewidth]{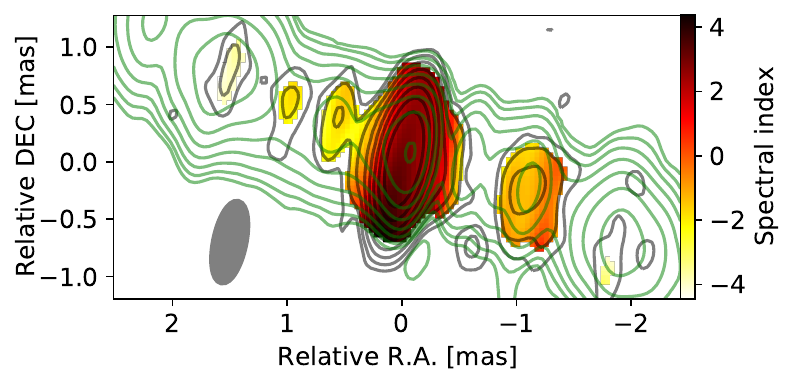}\\
    
    \includegraphics[width=0.49\linewidth]{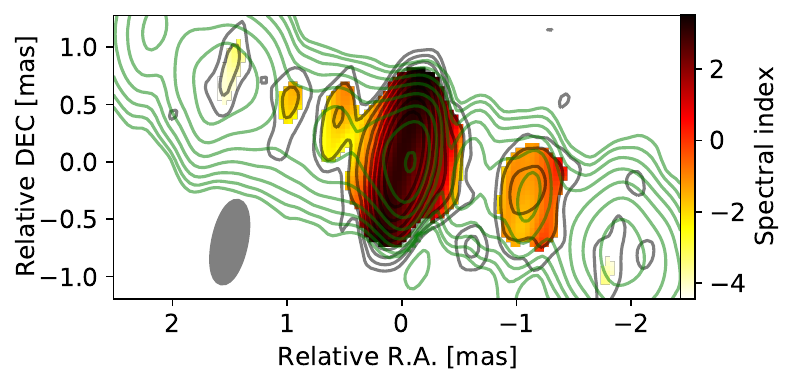}\hfill
    \includegraphics[width=0.49\linewidth]{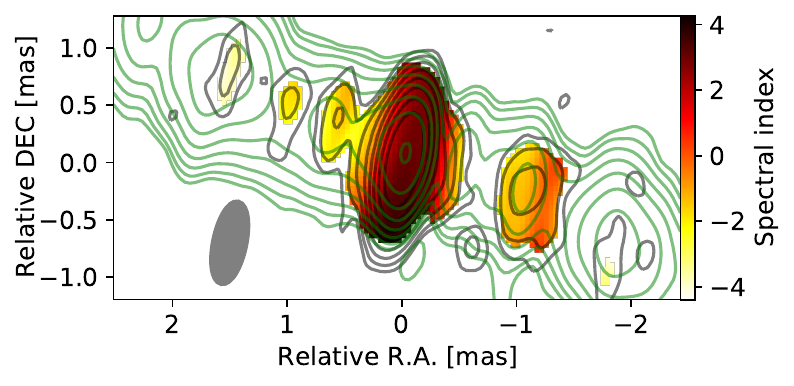}\\
    
    \includegraphics[width=0.49\linewidth]{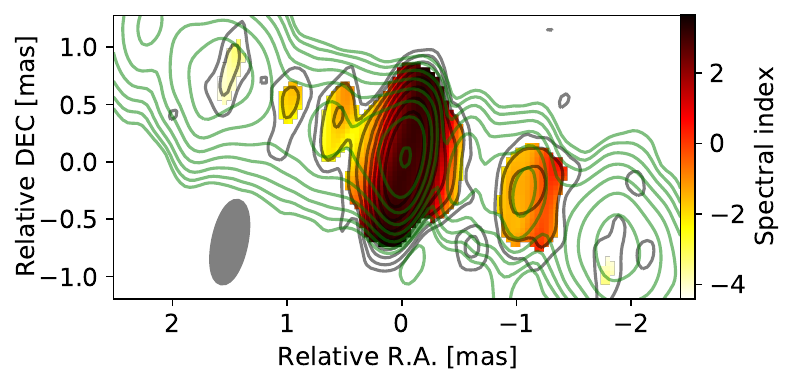}
    \includegraphics[width=0.49\linewidth]{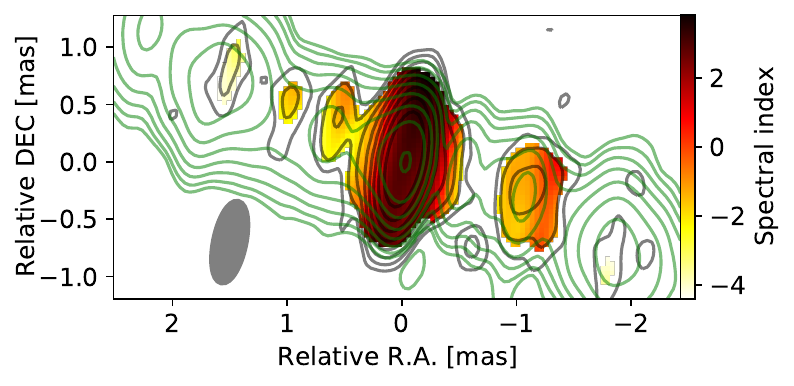}
    \caption{(43-86)\,GHz spectral index images with different shifts applied. Starting from the top-left panel the shifts applied are: $(x,y) = (2,1)$, $(x,y) = (2,0)$, $(x,y) = (2,-1)$, $(x,y) = (1,1)$, $(x,y) = (1,-1)$, $(x,y) = (0,1)$, $(x,y) = (0,0)$, $(x,y) = (0,-1)$.}
    \label{fig:othersm_align}
\end{figure*}

Secondly, in this section, we show in Fig.\,\ref{fig:othersm_align} the alternative (43-86)\,GHz spectral index maps obtained employing the map shown in the main text (Fig.\,\ref{fig:86GHzmap}) but with the different possible alignments discussed in Sect.\,\ref{sec:spectralindex}.

\begin{figure}[t]
    \centering
    \includegraphics[width=\linewidth]{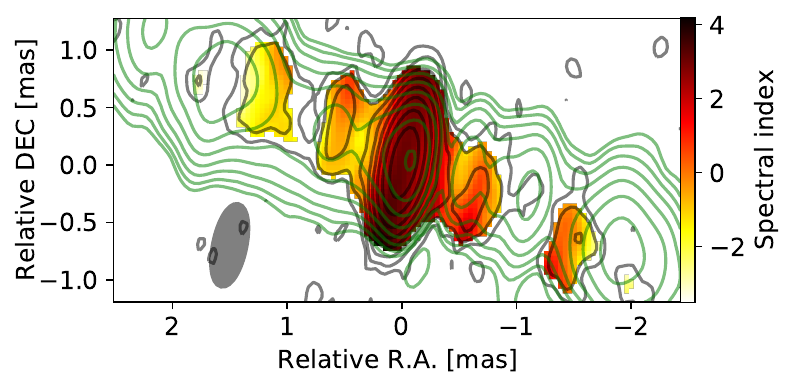}\par
    \caption{Alternative (43-86)\,GHz spectral index map using the map shown in Fig.\,\ref{fig:other86}. The black contours are for the 86\,GHz emission, while the green ones for the 43\,GHz image.}
    \label{fig:alternative}
\end{figure}

Finally, we show the alternative (43-86)\,GHz spectral index image (Fig.\,\ref{fig:alternative}) obtained using Fig.\,\ref{fig:other86}. 
The position angle in both jets at 86\,GHz appears to be slightly off, as indicated by the transverse gradient in the spectral index around 0.5 mas downstream of the core on both jets (we highlight how such a gradient persists with different core shifts applied).
Nonetheless, the core in the spectral index map remains highly inverted ($\alpha_\mathrm{thick} > 2.5$), consistent with Fig.\,\ref{fig:sm}.

\end{appendix}


\end{document}